\documentclass[apj, iop]{emulateapj}
\usepackage{natbib}
\usepackage{hyperref}

\newcommand{\NII}{[N{\footnotesize II}]}

\newcommand{\OIII}{[O{\footnotesize III}]} 
\newcommand{\SII}{[S{\footnotesize II}]}

\newcommand{\ha}{H$\alpha$}
\newcommand{\hbeta}{H$\beta$}

\newcommand{\msun}{$M_\odot$}
\newcommand{\mbh}{$M_{\rm BH}$}

\shorttitle{Local $M_{\rm BH} - M_{\star}$ Relations}
\shortauthors{Reines et al.}

\begin{document}

\title{Relations Between Central Black Hole Mass and Total Galaxy Stellar Mass in the Local Universe}

\author{Amy E. Reines\altaffilmark{1}}
\affil{Department of Astronomy, University of Michigan, 1085 South University Avenue, Ann Arbor, MI 48109, USA}
\email{reines@umich.edu}

\and

\author{Marta Volonteri}

\affil{Institut dÕAstrophysique de Paris, Sorbonne Universit\`{e}s, UPMC Univ Paris 6 et CNRS, UMR 7095, 98 bis bd Arago, 75014 Paris, France}

\altaffiltext{1}{Hubble Fellow}

\begin{abstract}

Scaling relations between central black hole (BH) mass and host galaxy properties are of fundamental importance to studies of BH and galaxy evolution throughout cosmic time.  Here we investigate the relationship between BH mass and host galaxy total stellar mass using a sample of 262 broad-line active galactic nuclei (AGN) in the nearby Universe ($z < 0.055$), as well as 79 galaxies with dynamical BH masses.  The vast majority of our AGN sample is constructed using Sloan Digital Sky Survey spectroscopy and searching for Seyfert-like narrow-line ratios and broad \ha\ emission.  BH masses are estimated using standard virial techniques.  We also include a small number of dwarf galaxies with total stellar masses $M_{\rm stellar} \lesssim 10^{9.5}~M_\odot$ and a sub-sample of the reverberation-mapped AGNs.  Total stellar masses of all 341 galaxies are calculated in the most consistent manner feasible using color-dependent mass-to-light ratios.  We find a clear correlation between BH mass and total stellar mass for the AGN host galaxies, with $M_{\rm BH} \propto M_{\rm stellar}$, similar to that of early-type galaxies with dynamically-detected BHs.  However, the relation defined by the AGNs has a normalization that is lower by more than an order of magnitude, with a BH-to-total stellar mass fraction of $M_{\rm BH}/M_{\rm stellar} \sim 0.025\%$ across the stellar mass range $10^8 \leq M_{\rm stellar}/M_\odot \leq 10^{12}$.  This result has significant implications for studies at high redshift and cosmological simulations in which stellar bulges cannot be resolved. 

\end{abstract}

\keywords{galaxies: active -- galaxies: evolution -- galaxies: nuclei -- galaxies: Seyfert}

\section{Introduction}

A growing body of evidence suggests that supermassive black hole (BH) masses scale with the large-scale properties of their host galaxies, primarily of the bulge component \citep[e.g., bulge mass, velocity dispersion, infrared luminosity, see for instance][and references therein]{Magorrian1998,gebhardtetal2000,fm00,MarconiHunt2003,Haring2004,Gultekin2009,2012MNRAS.419.2497B,2013ARAA..51..511K,McConnell2013}.  These correlations on the one hand hint to a joint evolution of BHs and galaxies, and contain crucial information on the cosmic assembly of structures; on the other hand they provide a way to estimate BH masses via a proxy which is often much easier to measure.  Extending these estimates to a statistical ensemble of galaxies allows one eventually to derive a census of BHs, such as their mass functions and total mass density locked into BHs \citep[e.g.,][]{Marconi2004,Shankar2004,Merloni2004,2009ApJ...690...20S,Gultekin2009,2012AdAst2012E...7K}. 

In the local Universe, $z=0$, benchmark BH masses are measured through direct methods, such as stellar and gas kinematics, and at the time of writing $\sim$90 galaxies have dynamical BH masses \citep[see][and http://blackhole.berkeley.edu]{2013ARAA..51..511K,McConnell2013}. The masses of the bulges
of the host galaxies can also be measured with good precision, using either photometric bulge/disk decomposition \citep[e.g.,][]{MarconiHunt2003} or kinematical fitting \citep{Haring2004} coupled with assumptions of the mass-to-light ratio \citep[see][for a discussion]{2013ARAA..51..511K}. 

\begin{figure*}[!t]
\includegraphics[width=7.2in]{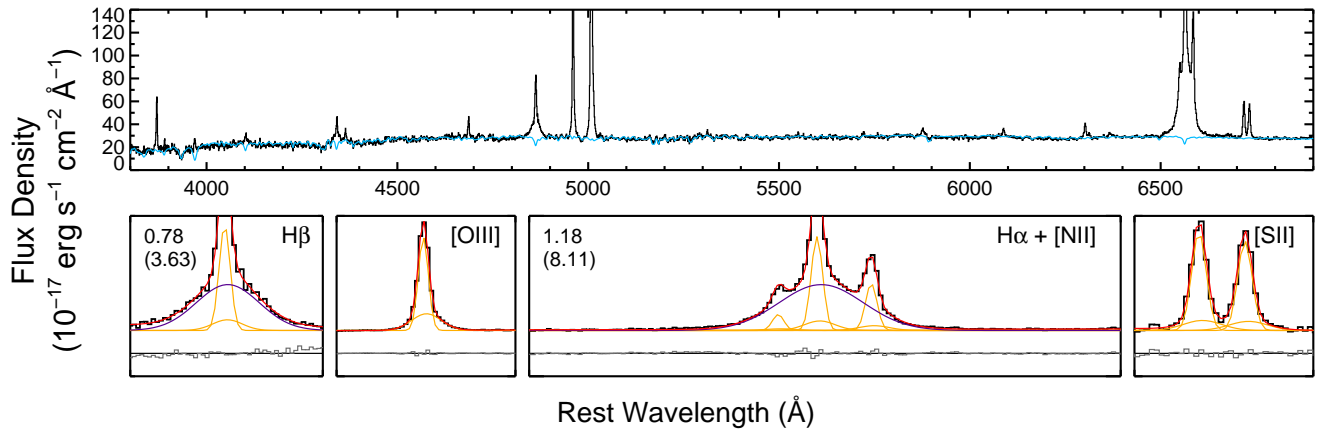}
\caption{\footnotesize Spectrum of a broad-line AGN illustrating our fitting method. Top: the redshift-corrected spectrum is shown in black and the continuum plus absorption-line model is plotted in blue.  Bottom: chunks of the emission-line spectrum (after subtracting the continuum and absorption-line model).  The best-fit models for the emission line regions are shown in red.  The individual narrow-line Gaussian components are plotted in yellow.  Broad \ha\ and H$\beta$ Gaussian components are plotted in dark blue.  The residuals are shown in gray with a vertical offset for clarity.  In the upper left-hand corner of the \ha\ and H$\beta$ chunks we show the reduced $\chi^2$ values.  The reduced $\chi^2$ values from the fits {\it not} including a broad component are shown below in parenthesis for comparison. 
\label{fig:spec}}
\end{figure*}

Estimates of the relative mass of BHs and their host galaxies at higher redshift, which are of fundamental importance to establish the timing of their growth, do not have access to the same wealth of information available locally. The BH masses are measured through indirect methods, and their uncertainties are discussed at length in the literature \citep[e.g.,][]{vestergaardpeterson2006,Shen2013}. The host properties are also estimated very differently. Except for gravitationally lensed galaxies \citep{2006ApJ...649..616P} and/or HST images \citep{2013ApJ...767...13S,2015ApJ...799..164P}, which cannot provide a large statistical sample, reliable decomposition into the bulge and disk components are very difficult due to lack of spatial resolution and sensitivity \citep[but see][]{2014MNRAS.445.1261S}. Normally, the total stellar mass is used instead, estimated through assuming a mass-to-light ratio or SED fitting \citep[e.g.,][]{Jahnke2009,2011ApJ...741L..11C,2012MNRAS.420.3621T,Merloni2010,2014MNRAS.443.2077B,2015ApJ...802...14S}, sometimes trying to select quasars with bulge-dominated host galaxies \citep{Decarli2010} to lessen the discrepancy between bulge mass and total stellar mass. The scaling between BH and total stellar mass at high-$z$ is then often compared to the scaling between BH and bulge mass at $z=0$.  \citet{laueretal2007} expose important biases incurred when investigating the potential evolution of BH scaling relations by comparing samples at different redshifts with different selection criteria (e.g., AGN activity vs.\ host galaxy properties).

In this paper we aim at quantifying on a local sample the difference between using total stellar mass and bulge stellar mass to calculate the BH-to-host relationship. While the tightest correlation appears to be the one between the BH and the bulge, we wish to provide a benchmark for high-redshift studies which cannot avail themselves of bulge masses (or dynamical BH masses).  We therefore investigate the relationship between BH mass and total stellar mass in a large sample of nearby ($z \sim 0$) broad-line AGNs using techniques for estimating BH and galaxy masses similar to those used at $z>0$.  Making use of active BHs has the added advantage of extending our sample to the lowest-mass BHs known in galaxy nuclei \citep{reinesetal2013,baldassareetal2015}.   

\section{Sample of Broad-line AGN}\label{sec:sample}

We construct our sample of broad-line AGNs by analyzing Sloan Digital Sky Survey (SDSS) spectra of $\sim$67,000 emission-line galaxies and searching for objects exhibiting broad \ha\ emission (signifying dense gas orbiting a massive BH) as well as narrow emission-line ratios indicating photoionization by an accreting massive BH.  Our parent sample of emission-line galaxies is culled from the NASA-Sloan Atlas (NSA), which is based on the SDSS Data Release 8 (DR8) spectroscopic catalog \citep{aiharaetal2011}.  While we use the NSA for selecting our parent sample of galaxies, we use our own software to analyze the SDSS spectra and search for broad-line AGN.  Distance estimates come from the {\tt zdist} parameter in the NSA, which is based on the SDSS NSA redshift and the peculiar velocity model of \citet{willicketal1997}.  We assume $H_0=70$ km s$^{-1}$ Mpc$^{-1}$.

\subsection{Parent Sample of Emission-line Galaxies}

The NSA catalog of nearby galaxies ($z \le 0.055$) provides a reanalysis of SDSS optical photometry using SDSS $ugriz$ images with the improved background-subtraction technique described in \citet{blantonetal2011}. The NSA also provides a reanalysis of spectroscopic data from the SDSS using the methods described in \citet{yanblanton2012} and \citet{yan2011}.  We select emission-line galaxies in the NSA by imposing modest signal-to-noise (S/N) cuts on emission-line measurements reported in the NSA.  We require S/N $\ge 3$ for the flux and S/N $>1$ for the equivalent width (EW) of the \ha, \NII$\lambda$6584, and \OIII$\lambda$5007 emission lines.  We also require S/N $\ge 2$ for \hbeta\ and the \SII$\lambda \lambda$6716,6731 doublet.  This leaves us with a parent sample of 66,945 galaxies.  

\begin{figure*}[!t]
$\begin{array}{cc}
{{\includegraphics[height=2.7in]{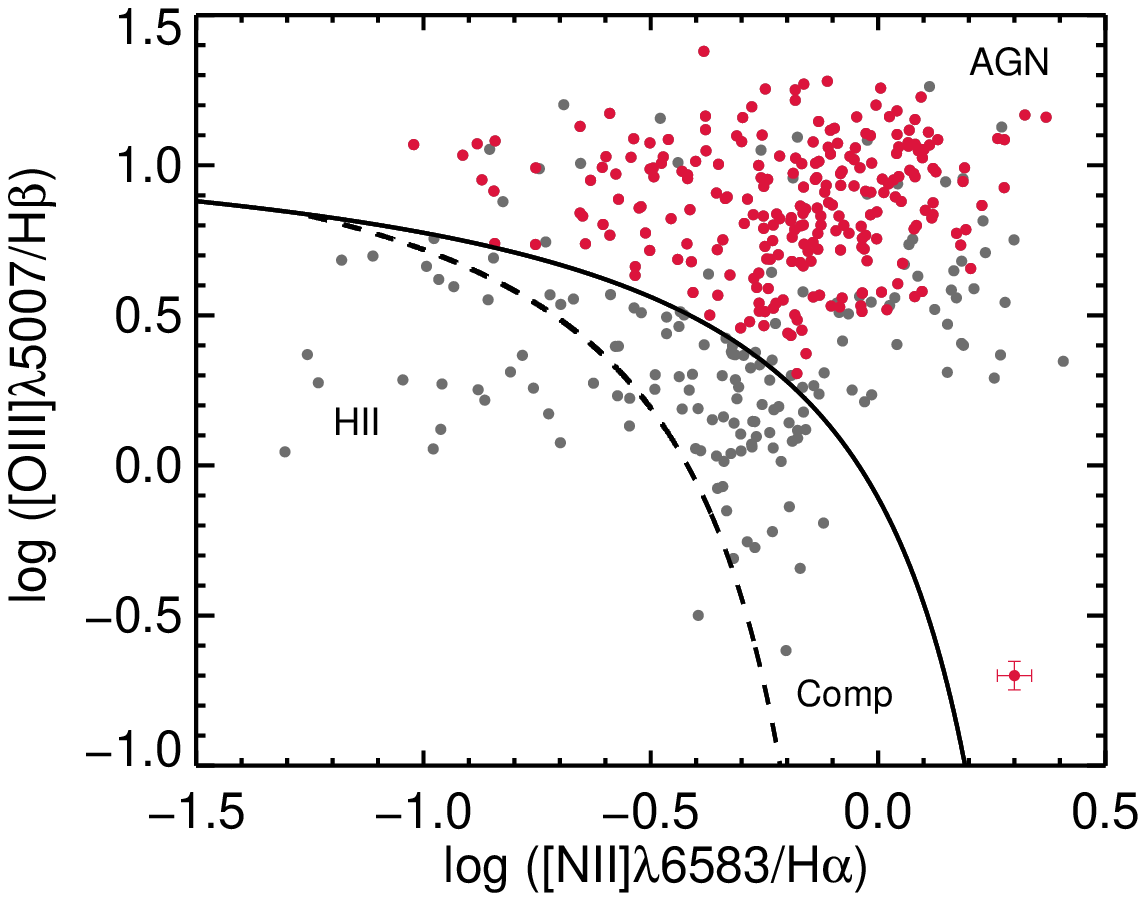}}} &
\hspace{-0.3cm}
{{\includegraphics[height=2.7in]{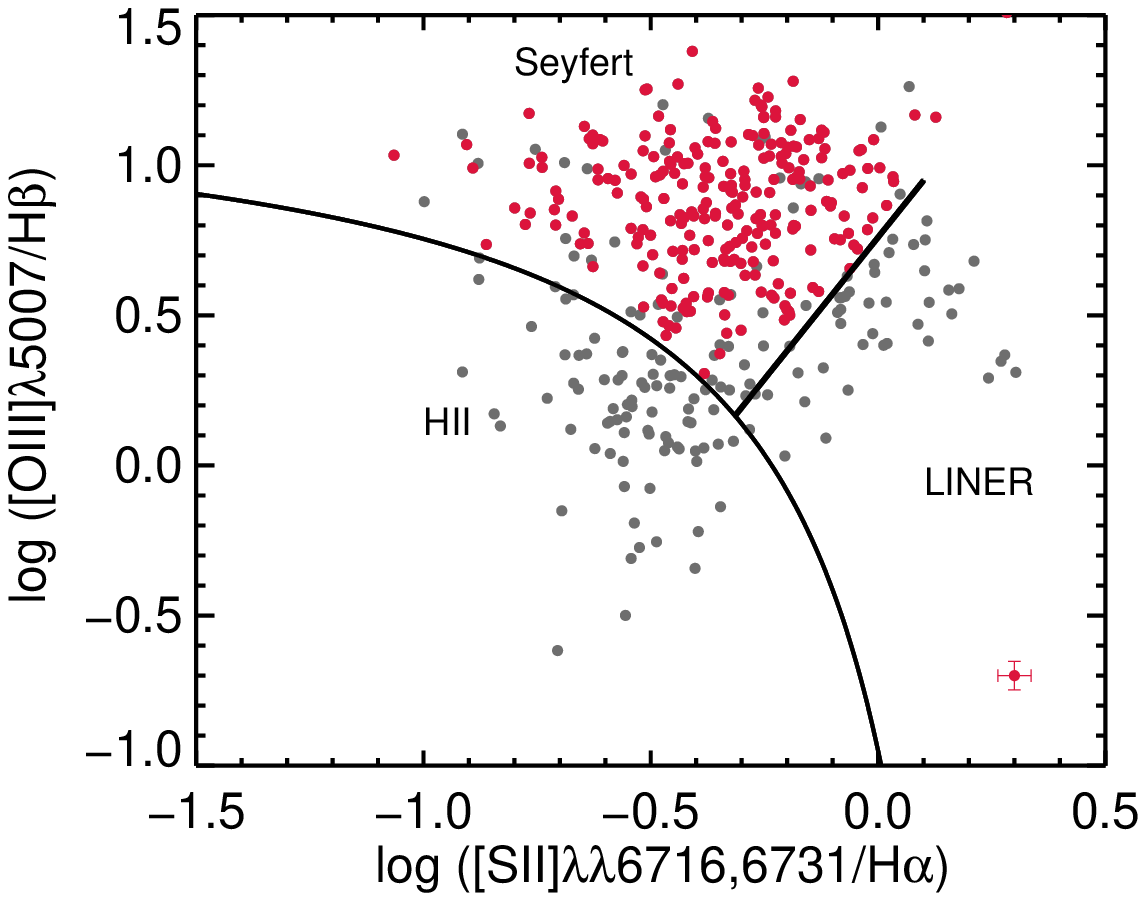}}}
\end{array}$
\caption{\footnotesize Narrow-line diagnostic diagrams for sources with detectable broad \ha\ emission.  We use the classification scheme outlined in \citet{kewleyetal2006}.  Our sample of broad-line AGNs is restricted to objects with narrow line ratios placing them in both the AGN region of the \OIII/\hbeta\ vs.\ \NII/\ha\ diagram and the Seyfert region of the \OIII/\hbeta\ vs.\ \SII/\ha\ diagram (red points).  The typical error for the red points is shown in the lower right corner of each plot.
\label{fig:bpt}}
\end{figure*}

\begin{deluxetable*}{ccccccccc}
\tabletypesize{\footnotesize}
\tablecaption{Sample of 244 Broad-line AGN}
\tablewidth{0pt}
\tablehead{
\colhead{ID} & \colhead{NSAID} & \colhead{SDSS Name} & \colhead{Plate-MJD-Fiber} &
\colhead{$z$dist}  & \colhead{$M_i$ (Host)} & \colhead{$g-i$ (Host)} & \colhead{log $M_\star$} & \colhead{log $M_{\rm BH}$}  \\
\colhead{(1)} & \colhead{(2)} & \colhead{(3)} & \colhead{(4)} & \colhead{(5)} & \colhead{(6)} & \colhead{(7)} & \colhead{(8)} & \colhead{(9)} }
\startdata
1 & 25955 & J000907.90+142755.8 & 752-52251-320 & 0.0422 & $-21.81$ & $1.06$ & 10.68 & 6.2 \\
2 & 22075 & J004236.86$-$104922.0 & 655-52162-58 & 0.0424 & $-21.10$ & $1.10$ & 10.43 & 7.1 \\
3 & 6452 & J012159.81$-$010224.3 & 398-51789-10 & 0.0548 & $-22.77$ & $0.49$ & 10.47 & 7.7 \\
4 & 23318 & J021011.49$-$090335.5 & 667-52163-506 & 0.0419 & $-22.60$ & $0.99$ & 10.92 & 8.1 \\
5 & 11183 & J024912.86$-$081525.7 & 456-51910-77 & 0.0296 & $-20.64$ & $0.96$ & 10.11 & 5.7
\enddata
\tablecomments{Column 1: identification number assigned in this paper. 
Column 2: NSA identification number. Column 3: SDSS name. Column 4: Plate-MJD-Fiber of analyzed spectra.
Column 5: {\tt zdist} parameter in the NSA, which is based on the SDSS NSA redshift and the peculiar velocity model of \citet{willicketal1997}.
Column 6: absolute $i$-band magnitude of the host galaxy.
Column 7: $g-i$ color of the host galaxy. Magnitudes and colors have been corrected for foreground Galactic extinction \citep{schlegeletal1998}
and the AGN contribution has been removed as described in \S\ref{sec:mstar}.
Column 6: log host galaxy stellar mass in units of $M_\odot$, corrected for AGN contribution.  Uncertainties are on the order of 0.3 dex.
Column 7: log black hole mass in units of $M_\odot$.  Uncertainties are on the order of 0.5 dex. \\ \\
(This table is available in its entirety in a machine-readable form in the online journal.  A portion is shown here for guidance regarding its form and content.)}
\label{tab:sample}
\end{deluxetable*}

\subsection{Spectral Analysis and Selection of Broad-line AGNs}\label{sec:agn}

We retrieved the SDSS spectra of our entire parent sample of galaxies and analyzed them with customized software that is described in detail in \citet{reinesetal2013} and briefly reviewed here for completeness.  First we model and remove the stellar continuum and absorption lines from the host galaxy using simple stellar population model templates spanning a range of ages and metallicities.  Next we model the narrow emission line profile based on the \SII$\lambda \lambda$6716,6731 doublet.  Once we have a suitable model of the \SII\ doublet, we use it as a template for fitting the narrow emission lines in the \ha\ + \NII$\lambda \lambda$6548,6583 complex.  We fit the \ha\ + \NII\ complex twice, first with the narrow lines only and then allowing a broad \ha\ component.  We accept the fit with the broad \ha\ component if statistically warranted (reduced $\chi^2$ is improved by more than 50\%\footnote{Reines et al.\ (2013) used a threshold of 20\% since they were focused on dwarf galaxies with low-mass BHs, which can have weak broad \ha\ emission.  Here we choose a higher threshold to help eliminate objects with marginally detected broad lines in higher mass galaxies.}) and the FWHM of the broad \ha\ component is $\ge 500$ km s$^{-1}$ after correcting for the fiber-dependent instrumental resolution (e.g., see Figure \ref{fig:spec}).  This FHWM requirement avoids severe contamination from intensely star forming galaxies with moderately broadened bases on \ha.  We also measure fluxes of \hbeta, \OIII\ $\lambda$5007, and the \SII\ doublet to place objects on standard narrow-line diagnostic diagrams \citep{baldwinetal1981,veilleuxosterbrock1987,kewleyetal2001,kewleyetal2006,kauffmannetal2003agn}.   

Figure \ref{fig:bpt} shows the \OIII/\hbeta\ vs.\ \NII/\ha\ and  \OIII/\hbeta\ vs.\ \SII/\ha\ narrow-line diagnostic diagrams for all 
objects with detectable broad \ha\ emission.  To minimize contamination from potential sources of broad \ha\ other than ionized gas orbiting a BH (e.g., supernovae in star-forming galaxies and shocks in LINERs), we restrict our sample of broad-line AGN to those sources falling in both the AGN region of the \OIII/\hbeta\ vs.\ \NII/\ha\ diagram and the Seyfert region of the \OIII/\hbeta\ vs.\ \SII/\ha\ diagram.  We also visually inspect each individual object and cut sources with poor spectral fits (e.g., due to complicated line profiles including double-peaked lines) that may lead to erroneous BH masses based on broad \ha\ (see below).  A handful of objects are  also excluded for reasons described in \S\ref{sec:mstar}, leaving us with a final sample of 244 broad-line AGN (Tables \ref{tab:sample} and \ref{tab:lines}).   

\begin{deluxetable*}{ccccccccccc}
\tabletypesize{\footnotesize}
\tablecaption{Sample of 244 Broad-line AGN: Emission Line Measurements}
\tablewidth{0pt}
\tablehead{
\colhead{ID} & \colhead{(H$\beta)_n$} & \colhead{(H$\beta)_b$} & \colhead{[O {\footnotesize III}]} & \colhead{[N {\footnotesize II}]} & \colhead{(H$\alpha)_n$} & \colhead{(H$\alpha)_b$} & \colhead{[N {\footnotesize II}]} & \colhead{[S {\footnotesize II}]} & \colhead{[S {\footnotesize II}]}  & \colhead{FWHM (H$\alpha)_b$} \\ 
\colhead{} & \colhead{} & \colhead{} & \colhead{$\lambda$5007} & \colhead{$\lambda$6548} & \colhead{} & \colhead{} & \colhead{$\lambda$6583} & \colhead{$\lambda$6716} & \colhead{$\lambda$6731}  & \colhead{} \\ 
\colhead{(1)} & \colhead{(2)} & \colhead{(3)} & \colhead{(4)} & \colhead{(5)} & \colhead{(6)} & \colhead{(7)} & \colhead{(8)} & \colhead{(9)} & \colhead{(10)} & \colhead{(11)} }
\startdata
1 & 245(13) & \nodata & 1659(77) & 301(13) & 1192(31) & 757(63) & 892(39) & 250(16) & 219(11) & 1501 \\
2 & 481(26) & 2343(147) & 4960(87) & 271(14) & 2022(103) & 11321(84) & 803(24) & 505(27) & 417(29) & 2121 \\
3 & 664(47) & 2073(289) & 6841(203) & 892(60) & 3563(237) & 16872(168) & 2640(116) & 627(53) & 612(42) & 3281 \\
4 & 494(33) & \nodata & 1876(84) & 761(49) & 1802(116) & 6309(715) & 2252(150) & 748(57) & 585(69) & 7720 \\
5 & 32(5) & 331(20) & 296(20) & 19(2) & 176(15) & 524(18) & 57(5) & 29(3) & 28(3) & 1081
\enddata
\tablecomments{Column 1: identification number assigned in this paper. Columns 2-10: emission line fluxes with units of $10^{-17}$ erg s$^{-1}$ cm$^{-2}$.
Errors are given in parenthesis. We have not applied an extinction correction. The subscripts $n$ and $b$ indicate the narrow and broad components of the line, respectively. A three-dot ellipsis indicates that no line was detected.
Column 11: FWHM of the broad H$\alpha$ component. \\ \\
(This table is available in its entirety in a machine-readable form in the online journal.  A portion is shown here for guidance regarding its form and content.)}
\label{tab:lines}
\end{deluxetable*}

\subsection{Black Hole Masses and Luminosities}\label{sec:mbh}

Single-epoch spectroscopic BH masses are routinely estimated for broad-line AGNs \citep[e.g.,][]{greeneho2007b,vestergaardosmer2009,schulzewisotzki2010}.  Under the assumption that the broad-line region (BLR) kinematics are dominated by the gravity of the BH, the BH mass is given by $M_{\rm BH} \propto R \Delta V^2/G$.  The average gas velocity is inferred from the width of a broad emission line (typically H$\beta$) and the radius of the BLR is estimated from the radius-luminosity relation defined by reverberation-mapped AGN \citep[e.g.,][]{kaspietal2005,bentzetal2013}.  The proportionality constant depends on the unknown geometry and orientation of the BLR.  While these parameters have been seen to vary from object to object \citep{kollatschny2003,bentzetal2009lamp,denneyetal2010,barthetal2011}, a single scaling factor is generally adopted from calibrating the ensemble of reverberation-based BH masses to the $M_{\rm BH}-\sigma_{\star}$ relation \citep[e.g.,][]{gebhardtetal2000b,ferrareseetal2001,nelsonetal2004,onkenetal2004,greeneho2006msig,parketal2012,grieretal2013,hokim2014}. 

\begin{figure*}[!t]
$\begin{array}{ccc}
\hspace{-0.5cm}
{{\includegraphics[height=2.25in]{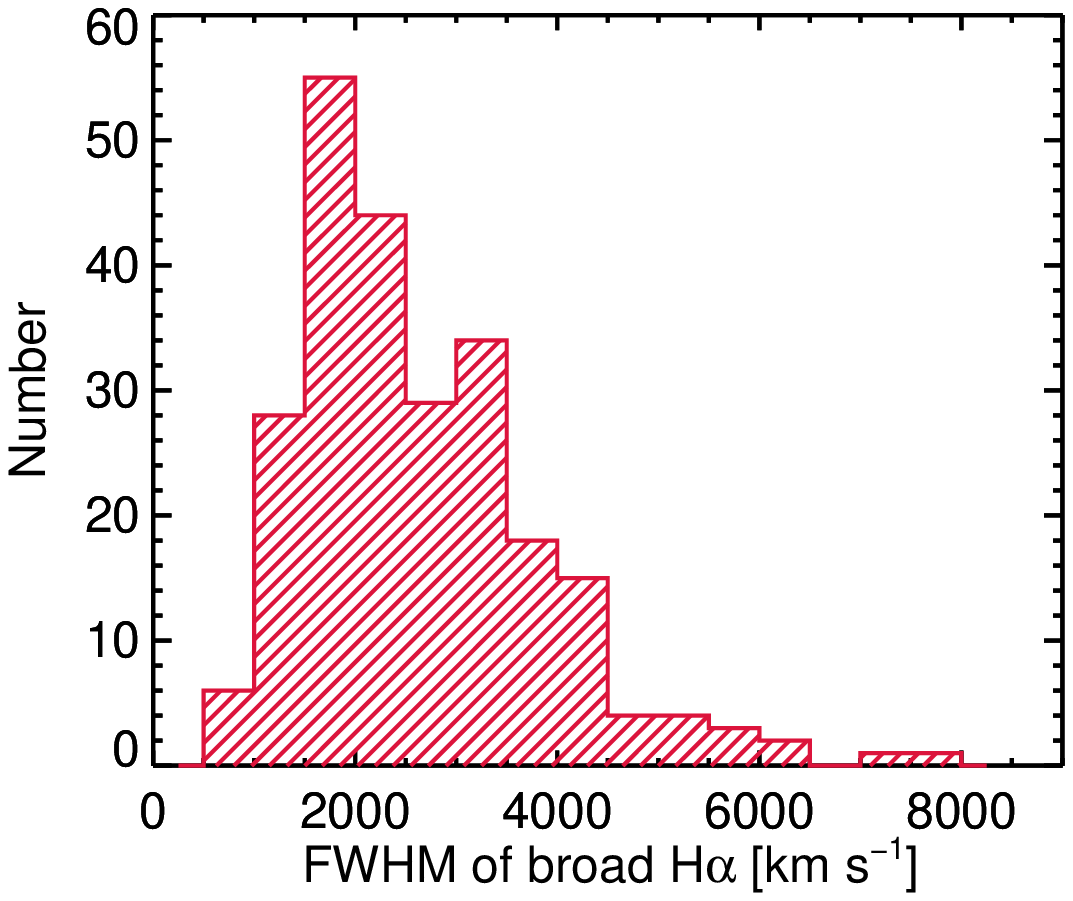}}} &
\hspace{-1.7cm}
{{\includegraphics[height=2.25in]{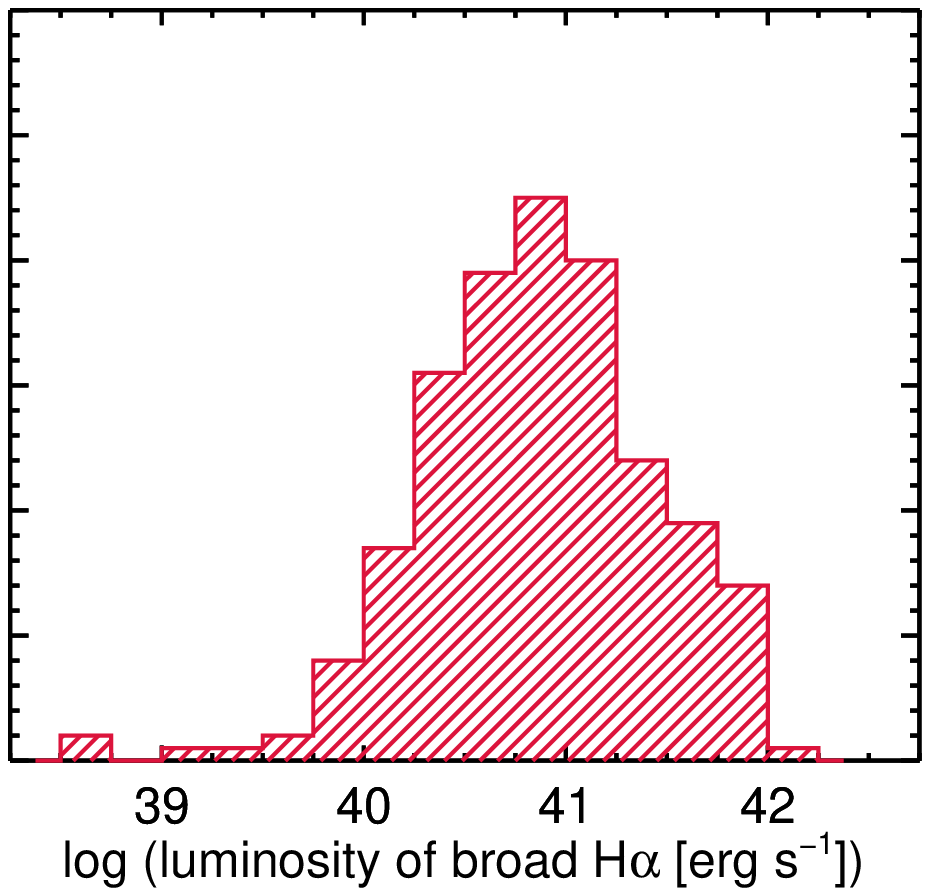}}} &
\hspace{-1.7cm}
{{\includegraphics[height=2.25in]{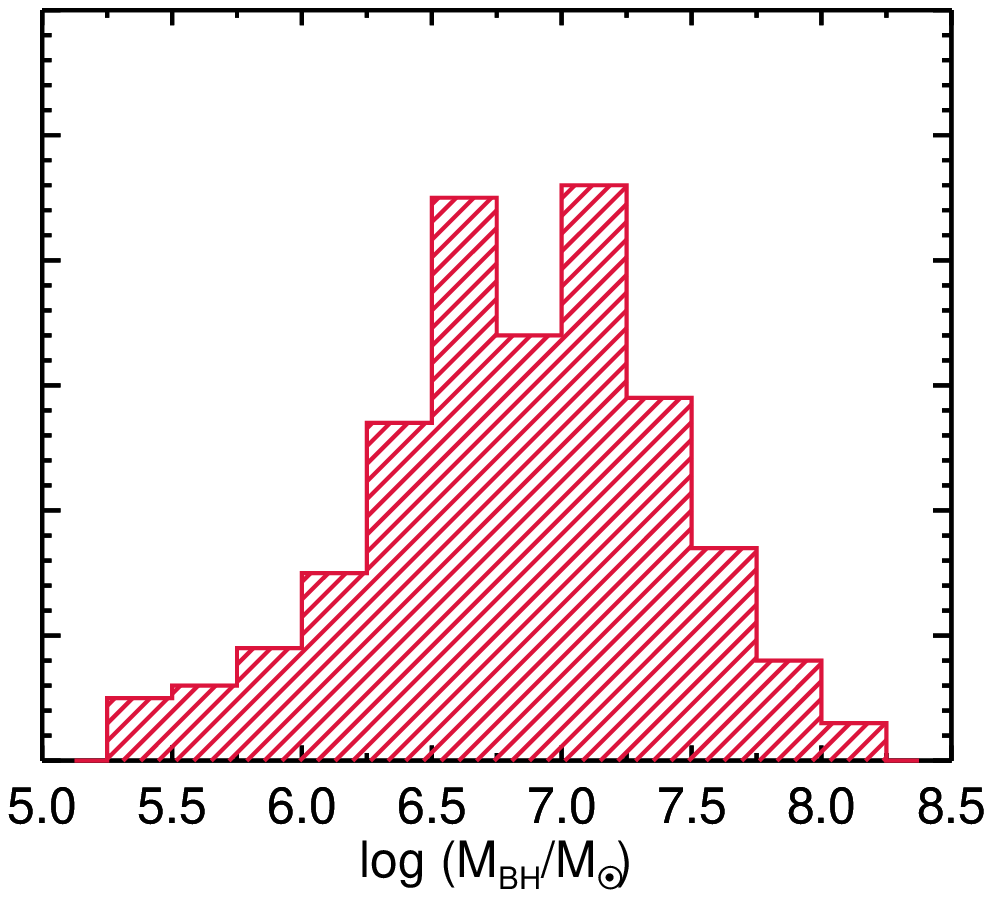}}}
\end{array}$
\caption{\footnotesize Distribution of the FWHM (left panel) and luminosity (middle panel) of broad \ha\ emission for our sample of nearby broad-line AGN.  The distribution of virial BH masses calculated from equation 1 is shown in the right panel. 
\label{fig:mbh}}
\end{figure*}

\begin{figure}[!t]
$\begin{array}{ccc}
{{\includegraphics[width=3.25in]{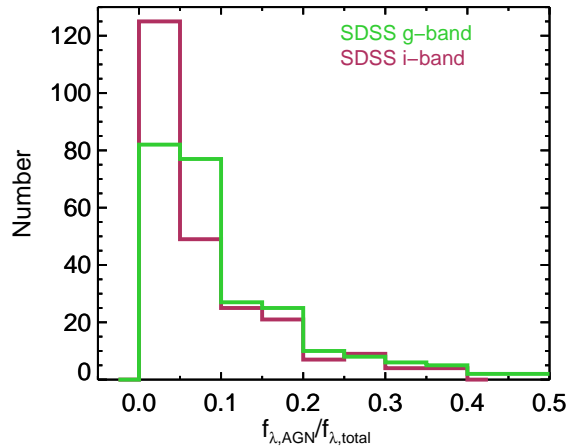}}} \\
{{\includegraphics[width=3.25in]{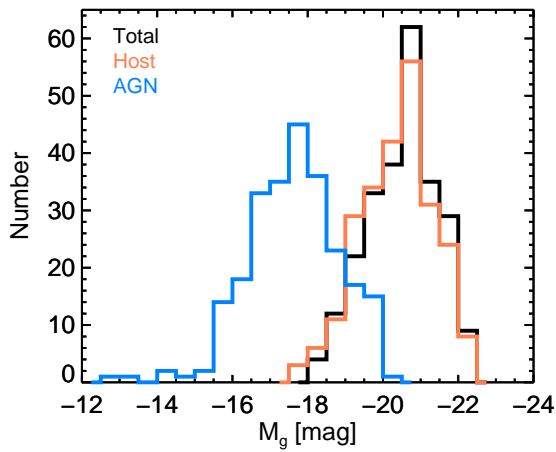}}} \\
{{\includegraphics[width=3.25in]{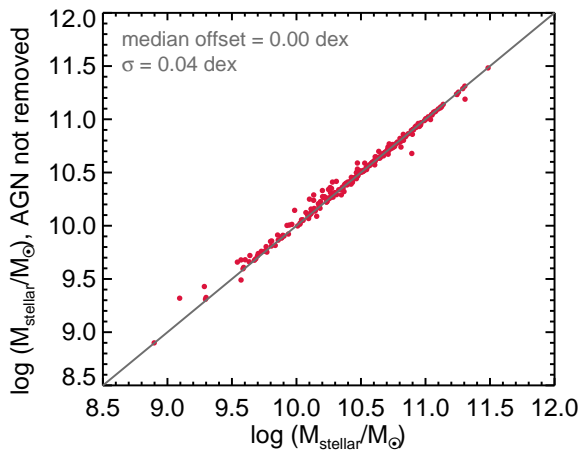}}}
\end{array}$
\caption{\footnotesize {\it Top:} Distributions of the ratio of AGN flux density to total (Host+AGN) flux density in the SDSS $g$ and $i$ bands for our sample of broad-line AGNs.  
{\it Middle:} Distributions of absolute $g$-band magnitudes for our sample of broad-line AGN.  The blue histogram shows AGN-only magnitudes, the orange histogram shows host galaxy-only magnitudes and the black histogram shows the total (Host+AGN) magnitudes. 
{\it Bottom:} Total stellar mass without correcting for AGN contamination versus total stellar mass corrected for AGN contamination.  See \S \ref{sec:mstar} for details. The line shows the one-to-one relation. 
\label{fig:agncontrib}}
\end{figure}

We estimate BH masses for our sample of broad-line AGN using the single-epoch virial mass estimator given by equation 5 in \citet{reinesetal2013}:

\begin{eqnarray}
{\rm log} \left({M_{\rm BH} \over M_\odot}\right) = {\rm log}~\epsilon + 6.57 + 0.47~{\rm log} \left({L_{\rm H_\alpha} \over 10^{42}~{\rm erg~s^{-1}} }\right) \\
\label{eqn:mbh}
\nonumber
+ 2.06~{\rm log} \left({\rm FWHM_{H\alpha} \over 10^{3}~{\rm km~s^{-1}} }\right).
\end{eqnarray}

\noindent
This equation was derived following the approach outlined in \citet{greeneho2005cal} for using the broad \ha\ line, but incorporates the updated radius-luminosity relationship from \citet{bentzetal2013}.  Here we adopt $\epsilon=1.075$, corresponding to the mean virial factor $<f>=4.3$ from \citet{grieretal2013}, where $\epsilon=f/4$ \citep[e.g.,][]{onkenetal2004}.  The distribution of BH masses for our sample of broad-line AGNs is shown in Figure \ref{fig:mbh}.  Viral BH mass estimates for broad-line AGNs are obviously very indirect and carry uncertainties of $\sim$0.5 dex \citep[e.g.,][]{Shen2013}.

We estimate the bolometric luminosities of the AGNs using the conversion between $L_{\rm H \alpha}$ and $L_{5100}$ given by Equation 1 in \citet{greeneho2005cal}, where $L_{\rm H \alpha}$ is the broad \ha\ luminosity and $L_{5100}$ is the continuum luminosity at 5100 \AA , and $L_{\rm bol} = 10.3 L_{5100}$ \citep{richardsetal2006}.  The range of bolometric luminosities is $41.5 \lesssim {\rm log}~L_{\rm bol} \lesssim 44.4$ and the median is log $L_{\rm bol} \sim 43.4$, approximately 2.5 dex larger than the median of the distribution of broad H$\alpha$ luminosities (see Figure \ref{fig:mbh}). 

\subsection{Total Stellar Masses of the Host Galaxies}\label{sec:mstar}

We estimate the total stellar masses, $M_{\rm stellar}$, of galaxies hosting broad-line AGN using mass-to-light ratios for $i$-band data ($M/L_i$) as a function of $g-i$ color following \citet{zibettietal2009}, after removing the AGN contribution to the integrated photometry.   

For each source, we estimate $g$-band and $i$-band flux densities of the AGN alone by constructing a mock AGN spectrum and convolving it with the SDSS filter throughput curves.  The mock AGN spectrum consists of a power law \citep[$f_\lambda \propto \lambda^\alpha$, where $\alpha = -1.56$ for $\lambda \le 5000$~\AA\ and $\alpha = -0.45$ for $\lambda > 5000$ \AA;][]{vandenberketal2001} plus the observed strong emission lines measured from the SDSS spectrum (\hbeta, \OIII, \ha, \NII, \SII).  We scale the mock AGN spectrum using the conversion between $L_{5100}$ and $L_{\rm H \alpha}$ \citep{greeneho2005cal}, where the broad \ha\ luminosity is measured from the SDSS spectrum (\S \ref{sec:agn}).  For the vast majority of our sample ($\sim$85\%), the AGN contribution to the total $g$-band and $i$-band flux densities is less than 20\% (see Figure \ref{fig:agncontrib}).  To minimize erroneous stellar mass estimates, we remove 5 sources from our sample in which the AGN dominates (i.e., contributes more than 50\% to) the integrated flux densities. 

\begin{figure}[!t]
\hspace{-.4cm}
\includegraphics[width=3.5in]{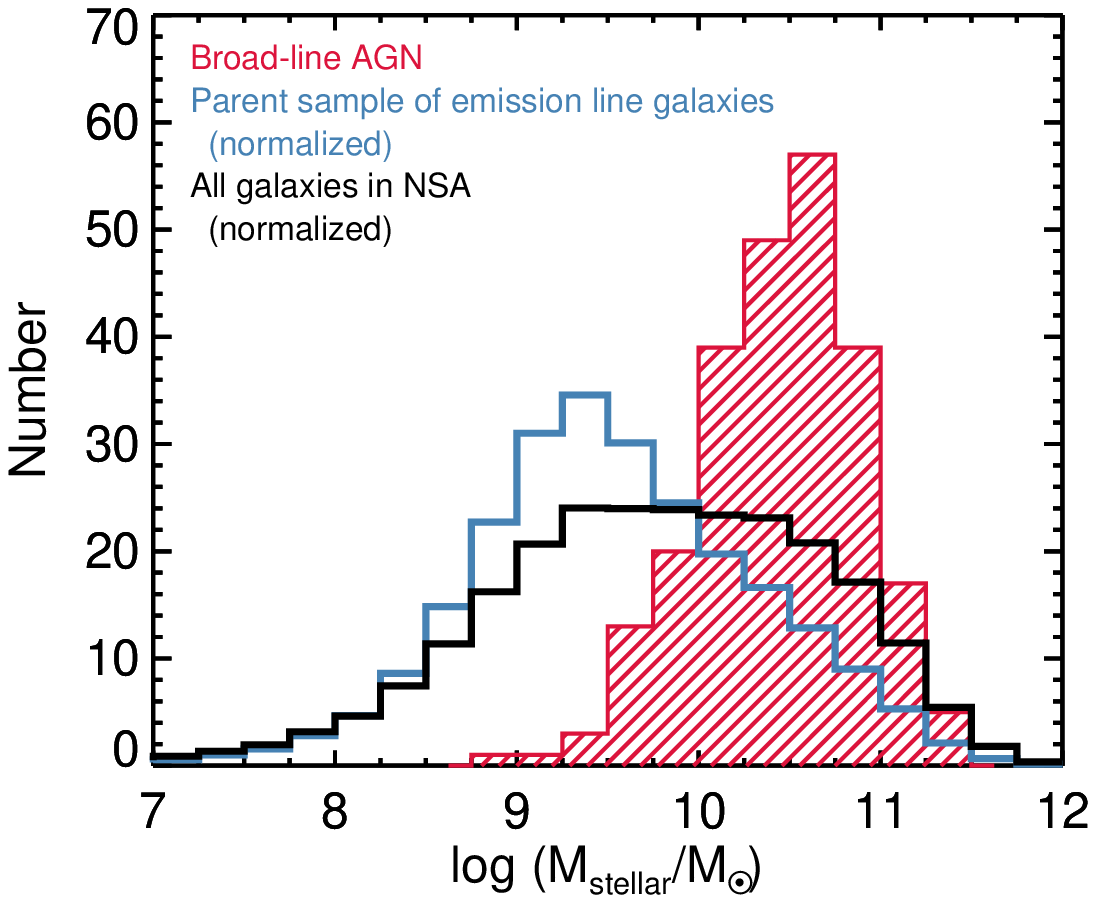}
\caption{\footnotesize The distribution of host galaxy total stellar masses for our sample of broad-line AGN (corrected for AGN contamination) is shown in red.  Our parent sample of emission-line galaxies is shown in blue, normalized to the number of galaxies in the red histogram (244 objects).  We also show the mass distribution for the full NSA catalog (no emission-line cuts), again normalized by the number of galaxies in the red histogram.  All masses were derived using $g$ and $i$-band data in the NSA with the color-dependent mass-to-light ratio given in equation \ref{eqn:mlratio}.    
\label{fig:mstar}}
\end{figure}

After removing the AGN contribution, the host-only AB magnitudes (corrected for Galactic reddening) are used to estimate galaxy stellar masses with a color-dependent mass-to-light ratio from \citet{zibettietal2009}:   

\begin{equation}
{\rm log}(M/L_i) = 1.032(g-i) - 0.963
\label{eqn:mlratio}
\end{equation}

\noindent
We adopt a solar absolute $i$-band magnitude of 4.56 mag \citep{belletal2003}.  Errors on the stellar masses are expected to be $\sim 0.3$ dex and are dominated by uncertainties in stellar evolution \citep{conroyetal2009}.      

\begin{figure}[!t]
$\begin{array}{cc}
\vspace{-0.3cm}
{{\includegraphics[width=3.25in]{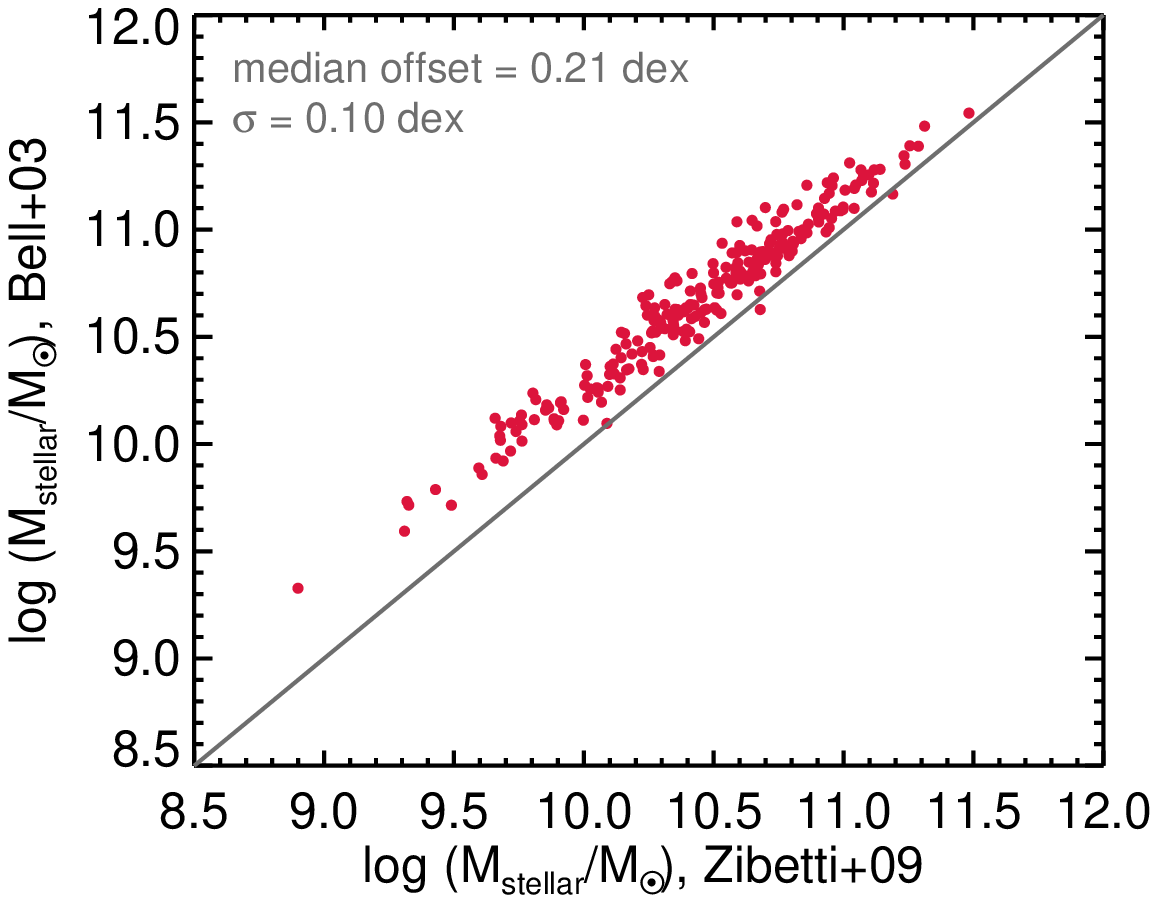}}} \\
{{\includegraphics[width=3.25in]{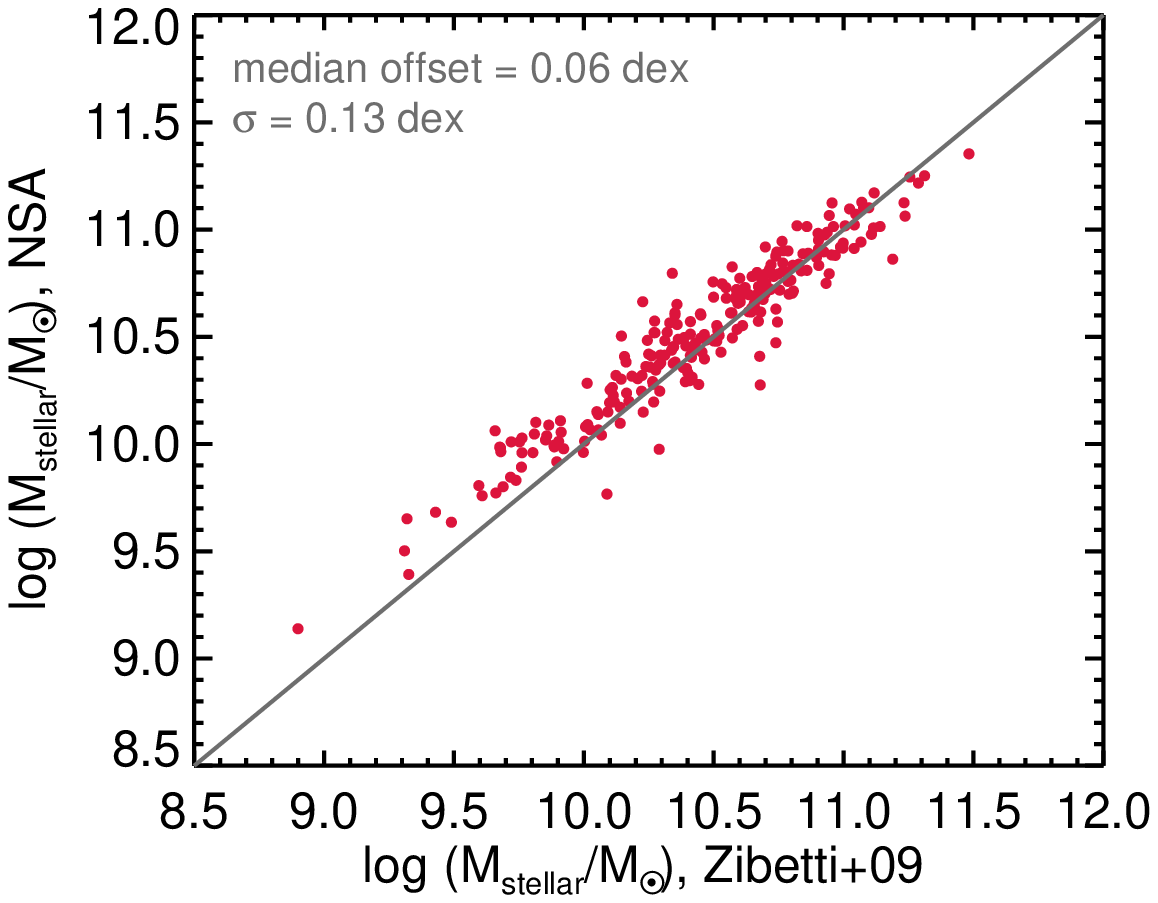}}}
\end{array}$
\caption{\footnotesize Comparison of stellar masses of our sample of broad-line AGN derived from different methods.  We calculate mass-to-light ratios for the SDSS $i$-band data as a function of $g-i$ color following both \citet{zibettietal2009} and \citet{belletal2003} and using integrated photometry provided in the NSA (AGN contribution has not been removed).  NSA masses are derived from the {\tt kcorrect} code, which is described in detail in \citet{blantonroweis2007}.  The lines show the one-to-one relation. 
\label{fig:masscomp}}
\end{figure}

In Figure \ref{fig:agncontrib} we compare total stellar masses of the host galaxies (with the AGN contribution removed) to those derived from the integrated photometry (without removing the AGN contribution).  The median offset is 0.00 dex with a $1\sigma$ scatter of 0.04 dex.  In some cases, the stellar mass actually increases slightly since the $g-i$ color gets redder once the (blue) AGN is removed, and a redder $g-i$ color increases $M/L_i$.  {The effect of correcting for AGN contamination is minimal since our sample is dominated by Seyferts of modest luminosity (\S\ref{sec:mbh}).}  The distribution of stellar masses is shown in Figure \ref{fig:mstar}. 

We also compare the stellar masses of our sample based on the mass-to-light ratios in \citet{zibettietal2009} to those based on mass-to-light ratios in \citet{belletal2003}, as well as the stellar masses provided in the NSA {based on SED fitting}.    For the \citet{belletal2003} masses, we again use $M/L_i$ as a function of $g-i$ color.  We then scale log $M/L_i$ down by $-0.093$ dex \citep{zibettietal2009,gallazzietal2008} to account for the differences between the Chabrier initial mass function (IMF) used in \citet{zibettietal2009} and the scaled Salpeter IMF used in \citet{belletal2003}.  {The NSA stellar masses are derived from the {\tt kcorrect} code \citep{blantonroweis2007}, which assumes a Chabrier IMF and fits broadband optical fluxes from the SDSS and ultraviolet fluxes from the {\it Galaxy Evolution Explorer} when available.}  Stellar masses from the three methods (using integrated photometry in the NSA) are compared in Figure \ref{fig:masscomp}.  The \citet{belletal2003} masses are systematically higher than the \citet{zibettietal2009} masses with a median offset of 0.21 dex ($\sigma$ = 0.10 dex) across the sample, with the largest discrepancies at low masses.  The NSA ({\tt kcorrect}) masses are more consistent with the \citet{zibettietal2009} masses with a median offset of 0.06 dex ($\sigma$ = 0.13 dex).

\section{Additional Objects}\label{sec:add}

We include the following additional objects in our investigation of the relationship between BH mass and host galaxy total stellar mass.

\subsection{Dwarf Galaxies Hosting Broad-line AGN}\label{sec:dwarfs}

\citet{reinesetal2013} carried out the first systematic search for AGNs in dwarf galaxies.  The vast majority of objects in that sample are narrow-line AGNs and composites as defined by \citet{kewleyetal2006}, for which we have no estimates of the BH masses.  However, 10 out of 136 AGN and composite objects have detectable broad \ha\ emission and narrow-line signatures suggesting the presence of an active BH (6 AGN and 4 composites).  This subsample includes the well-studied dwarf disk galaxy NGC 4395 hosting a Seyfert 1 nucleus \citep{filippenkosargent1989,filippenkoho2003}, two objects from the \citet{greeneho2007} sample of low-mass BHs, and the dwarf disk galaxy presented in \citet{dongetal2007}. \citet{reinesetal2013} also provide information on 15 galaxies with broad \ha\ emission, but narrow-line ratios dominated by star formation.  As described in that work, there is likely significant contamination from luminous Type II supernovae in that subsample and we therefore do not include those objects here.  This issue will be further addressed in a forthcoming paper (V.F.\ Baldassare et al., in preparation).  BH masses and total stellar masses for the 10 broad-line AGN and composites from \citet{reinesetal2013} are recomputed\footnote{Here we use $\epsilon=1.075$ in equation 1, rather than $\epsilon=1$ as in \citet{reinesetal2013}.  Stellar masses are computed using equation \ref{eqn:mlratio} rather than taken from the NSA as in \citet{reinesetal2013}.} here in the same way as our full sample of broad-line AGN (see \S\ref{sec:mbh} and \S\ref{sec:mstar}) and listed in Table \ref{tab:add}.

We note that 3 of the broad-line objects from \citet{reinesetal2013} are also included in our main sample of broad-line AGN in this work (including NGC 4395).  The remaining 7 objects are not recovered due to different selection criteria.  In this work we exclude composite objects and also impose a more stringent threshold for flagging a source as having broad \ha\ emission (see \S\ref{sec:agn}), since we are more concerned with having a clean sample than finding rare objects.  In contrast, \citet{reinesetal2013} focused on finding low-mass BHs that can have weak broad \ha\ emission in dwarf galaxies that tend to have more active star formation.  

Follow-up high resolution spectroscopy of the \citet{reinesetal2013} sample has led to the discovery of a new broad-line object with $M_{\rm BH} \sim 50,000~M_\odot$ \citep{baldassareetal2015}.  This object, designated RGG 118 (object ID 118 in the Reines et al.\ paper), has the smallest BH reported in a galaxy nucleus.  The BH mass for RGG 118 was estimated using equation 1 and here we estimate the stellar mass of RGG 118 to be  $M_{\rm stellar} \sim 2.7 \times 10^9~M_\odot$ using the SDSS $i$ and $g$-band photometry in the NSA with equation \ref{eqn:mlratio}. 

We also include the well-studied dwarf Seyfert 1 galaxy Pox 52 \citep{barthetal2004,thorntonetal2008}.  The mass of the BH in Pox 52 is $M_{\rm BH} \sim 3 \times 10^5~M_\odot$ \citep{thorntonetal2008}.  As Pox 52 is not in the SDSS footprint, we estimate the galaxy stellar mass using $B,V {\rm and~} K$-band photometry (corrected for the AGN contribution) provided by \citet{barthetal2004} and \citet{thorntonetal2008} with the following color-dependent mass-to-light ratio from \citet{zibettietal2009}:

\begin{equation}
{\rm log}(M/L_K) = 1.176(B-V) - 1.390.
\label{eqn:mlratio_K}
\end{equation}
 
\noindent
We adopt a solar absolute $K$-band magnitude of 3.32 mag \citep{belletal2003}.
The resulting stellar mass for Pox 52 is $M_{\rm stellar} \sim 4.3 \times 10^8~M_\odot$.  This is $\sim$2.8 times smaller than the stellar mass estimated by \citet{thorntonetal2008} using the {\tt kcorrect} code of \citet{blantonroweis2007}.  We adopt the \citet{zibettietal2009} mass for consistency with the rest of our sample.  

As with our main sample of broad-line AGNs (\S\ref{sec:sample}), we adopt uncertainties of 0.3 dex in stellar mass and 0.5 dex in BH mass for all of the dwarf galaxies discussed above.  However we caution that the virial BH masses derived for these objects are based on an extrapolation from more massive and luminous AGNs and may carry additional errors.  We do not include the dwarf galaxies Henize 2-10 \citep{reinesetal2011,reinesdeller2012}, Mrk 709 \citep{reinesetal2014} or J1329+3234 \citep{secrestetal2015} since the BH masses in these systems are uncertain by at least an order of magnitude.

It is worth noting here that optical searches for AGN in dwarf galaxies suffer from severe selection effects.  First, low-metallicity AGN, which are likely to reside in low-mass galaxies, have line ratios significantly different from Seyferts in more metal-rich systems and overlap with low-metallicity starbursts in the \OIII/\hbeta\ vs.\ \NII/\ha\ (i.e., BPT) diagram \citep{grovesetal2006,kewleyetal2013,reinesetal2014}.  Moreover, broad \ha\ from small accreting BHs can be very weak and difficult to detect.  Given the sensitivity of the SDSS spectra and our search volume ($z < 0.055$), \citet{reinesetal2013} estimate a minimum detectable BH mass of $M_{\rm BH} \sim 10^5~M_\odot$ if the BH is radiating at $\sim$10\% of its Eddington limit.  

\begin{deluxetable}{lrr}
\tabletypesize{\footnotesize}
\tablecaption{Additional Objects}
\tablewidth{0pt}
\tablehead{
\colhead{Name} & \colhead{log $M_\star$} & \colhead{log $M_{\rm BH}$}  \\
\colhead{(1)} & \colhead{(2)} & \colhead{(3)} }
\startdata
\cutinhead{Dwarf Galaxies with Broad-line AGN}
RGG 1 & 9.30 & 5.44(0.50) \\
RGG 9 & 9.24 & 5.00(0.50) \\
RGG 11 & 9.30 & 5.29(0.50) \\
RGG 20 & 9.29 & 6.10(0.50) \\
RGG 21 & 9.45 & 5.80(0.50) \\
RGG 32 & 8.90 & 5.28(0.50) \\
RGG 48 & 8.96 & 5.18(0.50) \\
RGG 118\tablenotemark{a} & 9.43 & 4.70(0.50) \\
RGG 119 & 9.12 & 5.42(0.50) \\
RGG 123 & 9.36 & 5.79(0.50) \\
RGG 127 & 9.36 & 5.21(0.50) \\
Pox 52\tablenotemark{b} & 8.63 & 5.48(0.50) \\
\cutinhead{Reverberation-Mapped AGN}
Mrk 590 & 11.38 & 7.57(0.07) \\
Mrk 79 & 10.41 & 7.61(0.12) \\
Mrk 110 & 10.14 & 7.29(0.10) \\
NGC 3227 & 10.36 & 6.77(0.10) \\
SBS 1116+583A & 10.00 & 6.56(0.08) \\
Arp 151 & 9.90 & 6.67(0.05) \\
Mrk 1310 & 9.62 & 6.21(0.08) \\
NGC 4051 & 10.16 & 6.13(0.14) \\
Mrk 202 & 9.92 & 6.13(0.17) \\
NGC 4253 & 10.34 & 6.82(0.05) \\
NGC 4395 & 8.90 & 5.45(0.14) \\
NGC 5273 & 10.25 & 6.66(0.16) \\
NGC 5548 & 10.79 & 7.72(0.02) \\
Mrk 817 & 9.87 & 7.59(0.07) \\
Mrk 290 & 9.52 & 7.28(0.06) \\
\cutinhead{Galaxies with Dynamical BH Masses}
M 32 & 8.77 & 6.39(0.18) \\
NGC 1316 & 11.48 & 8.23(0.07) \\
NGC 1332 & 10.92 & 9.17(0.06) \\
NGC 1374 & 10.33 & 8.77(0.04) \\
NGC 1399 & 11.17 & 8.94(0.33) \\
NGC 1407 & 11.43 & 9.67(0.05) \\
NGC 1550 & 11.02 & 9.59(0.07) \\
NGC 2960 & 10.72 & 7.03(0.02) \\
NGC 3091 & 11.29 & 9.57(0.04) \\
NGC 3377 & 10.14 & 8.25(0.23)
\enddata
\tablecomments{Column 1: object name.  Dwarf galaxies with the designation RGG are from \citet{reinesetal2013}.  
Column 2: log host galaxy stellar mass in units of $M_\odot$.  The AGN contribution has been removed for the dwarf galaxies and reverberation-mapped AGN.
All stellar masses are estimated using color-dependent mass-to-light ratios from \citet{zibettietal2009}. Uncertainties are on the order of 0.3 dex.
Column 3: log black hole mass in units of $M_\odot$.  
BH masses for the reverberation-mapped AGN are taken from the AGN Black Hole Mass Database \citep{bentzkatz2015}.  Dynamical BH masses
are taken from \citet{2013ARAA..51..511K}. 
\tablenotetext{a}{BH mass from Baldassare et al.\ (2015).}
\tablenotetext{b}{BH mass from \citet{thorntonetal2008}.}  \\ \\
(This table is available in its entirety in a machine-readable form in the online journal.  A portion is shown here for guidance regarding its form and content.)}
\label{tab:add}
\end{deluxetable}

\subsection{Reverberation-Mapped AGN}\label{sec:rm}

The most reliable AGN BH masses come from reverberation mapping \citep[e.g.,][]{petersonetal2004,bentzetal2009lamp,denneyetal2010,barthetal2011}.  Determining the time lag between the continuum flux and broad emission line variability gives the light travel time across the BLR, and in turn the BLR radius when multiplied by the speed of light.  The BLR radius-luminosity correlation derived from the sample of reverberation-mapped AGN ($\sim$50 objects) makes single-epoch virial BH masses for AGNs, such as the ones used in this work, possible \citep[e.g.,][]{kaspietal2005,bentzetal2013}.  The reverberation-mapped AGN also provide a link between single-epoch spectroscopic BH masses and dynamical BH masses, as the ensemble of reverberation-mapped BH masses is calibrated to the $M_{\rm BH}-\sigma_{\star}$ relation \citep[e.g.,][]{gebhardtetal2000b,ferrareseetal2001,nelsonetal2004,onkenetal2004,greeneho2006msig,parketal2012,grieretal2013,hokim2014}. 

In this work we include 15 reverberation-mapped AGNs with BH masses provided in the AGN Black Hole Mass Database \citep{bentzkatz2015}.  {We adopt BH masses calculated with a mean virial factor of $<f>=4.3$ \citep{grieretal2013} as we did for the single-epoch spectroscopic masses given by equation 1.}  We first cross-match the AGN Black Hole Mass Database with the NSA and find 19 matches.  
We require that galaxies have photometry in the NSA so we can calculate the host galaxy stellar masses consistently with the rest of our AGN sample (equation \ref{eqn:mlratio}).  We correct for the AGN contribution to the integrated photometry, assuming the same power law shape described in section \ref{sec:mstar}.  However, for most of the reverberation-mapped AGN, the normalization comes directly from $L_{5100}$ provided in the AGN Black Hole Mass Database.  There are a few cases where $L_{5100}$ is not available, yet we have a measurement of broad \ha\ from the SDSS spectrum.  For these, we normalize the AGN continuum as described in section \ref{sec:mstar}.  Emission lines are not included in the mock AGN spectrum for the reverberation-mapped AGN, as these measurements are not readily available in many cases.  We exclude 4 objects (from the initial 19 matches) in which the AGN dominates the integrated photometry to minimize unreliable stellar mass estimates.  BH masses and total stellar masses for the 15 reverberation-mapped AGNs used in this work are listed in Table \ref{tab:add}.  

{We note that six of the reverberation-mapped AGNs in Table \ref{tab:add} are included in our main sample of broad-line AGNs with SDSS spectroscopy.   As a consistency check, we compare the reverberation-mapped BH masses and those based on broad \ha\ emission measured from the SDSS spectra (equation 1).  For this limited sample of six objects, the spectroscopic BH masses are on average $\sim 0.4$ dex larger than the reverberation masses.}  

\subsection{Galaxies with Dynamically Detected BHs}\label{sec:KHmasses}

Benchmark BH masses come from dynamical methods, which rely on observations that spatially resolve the BH sphere of influence.  \citet{2013ARAA..51..511K} provide an inventory of BH mass measurements based on stellar dynamics, ionized gas dynamics, CO molecular gas disk dynamics and maser disk dynamics.  

We estimate total stellar masses of galaxies with dynamical BH mass measurements using the total absolute $K$-band magnitudes and $B-V$ colors provided by \citet{2013ARAA..51..511K} in their tables 2 and 3 with the color-dependent mass-to-light ratio from \citet{zibettietal2009} given in equation \ref{eqn:mlratio_K} above.  The results are listed in Table \ref{tab:add}. We include all objects summarized in \citet{2013ARAA..51..511K} except those with BH mass upper limits (2 elliptical galaxies and 2 spiral galaxies with pseudobulges), and galaxies without provided $B-V$ colors that are necessary for estimating the galaxy stellar masses (2 elliptical galaxies\footnote{Two additional elliptical galaxies do not have $B-V$ colors provided by \citet{2013ARAA..51..511K}, however they do have $M_{\rm bulge}$ which is equivalent to total stellar mass since these are ellipticals.  We include these galaxies in our sample and adopt a total stellar mass log~$M_{\rm stellar} =$ log~$M_{\rm bulge} - 0.33$, where the offset accounts for differences our assumed mass-to-light ratios.} and 3 spiral galaxies with pseudobulges).  

\begin{figure}[!t]
\includegraphics[width=3.3in]{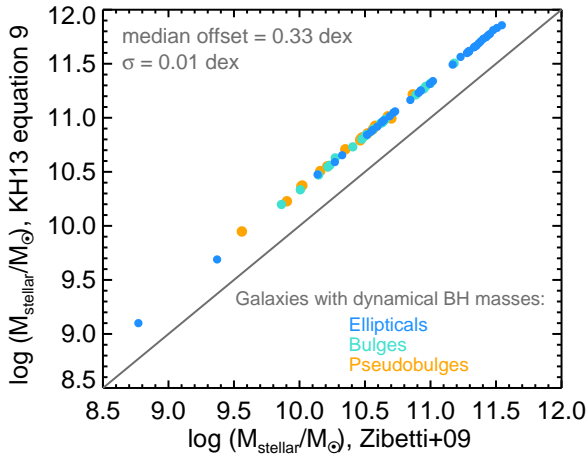}
\caption{\footnotesize Total stellar masses of galaxies with dynamically measured BH masses derived using $K$-band mass-to-light ratios as a function of $B-V$ color from  equation 9 in \citet{2013ARAA..51..511K} versus equation \ref{eqn:mlratio_K} here taken from \citet{zibettietal2009}.  
\label{fig:KHmasses}}
\end{figure}

\begin{figure*}[!t]
\hspace{1.75cm}
{\includegraphics[width=5.3in]{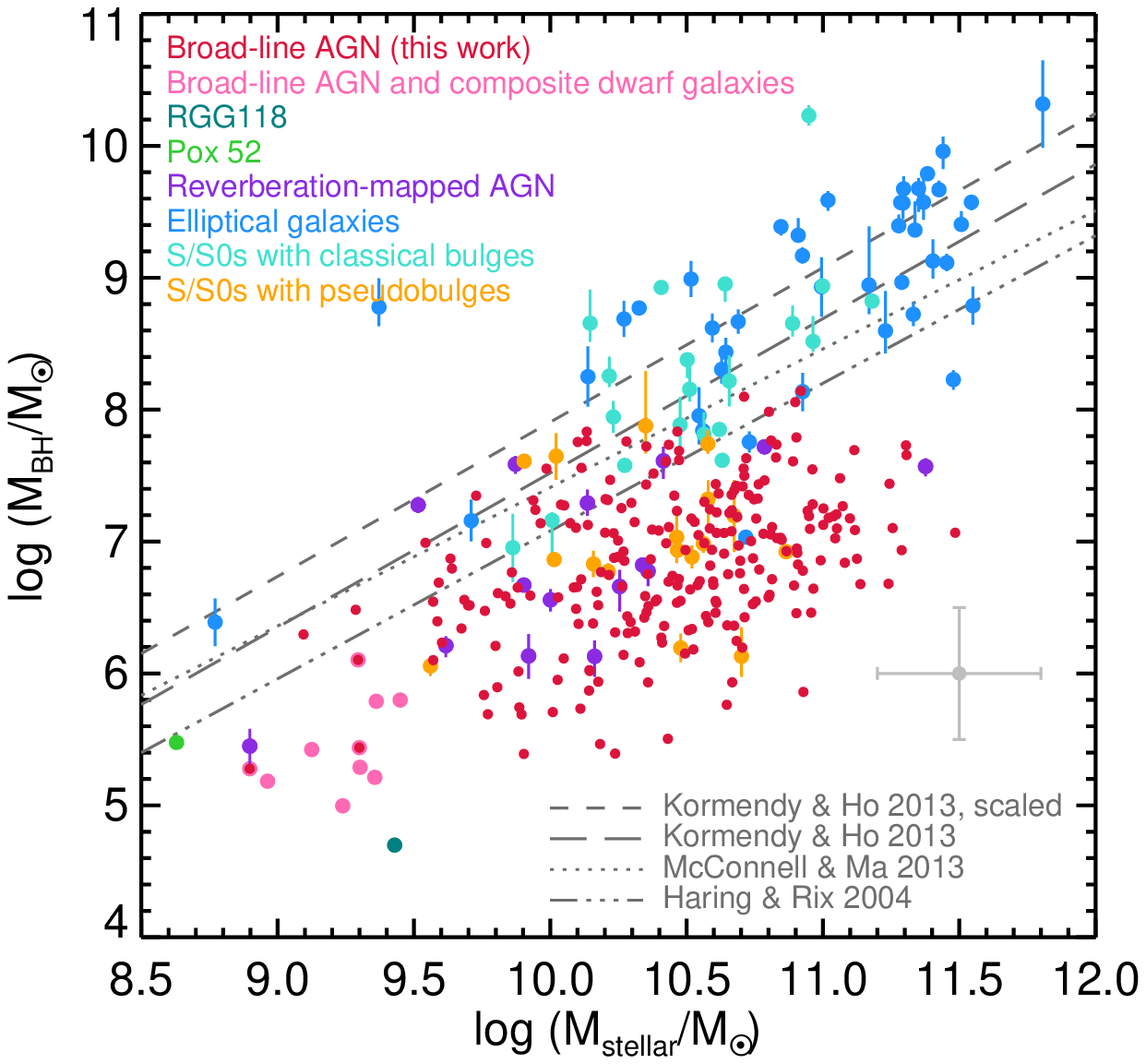}}
\caption{\footnotesize Left: Black hole mass versus total host galaxy stellar mass.  All stellar masses are estimated using color dependent mass-to-light ratios presented in \citet{zibettietal2009} (see \S\ref{sec:mstar} and \S\ref{sec:add}).  Our sample of 244 broad-line AGN for which we estimate virial BH masses from equation 1 are shown as red points.  The 10 broad-line AGN and composite dwarf galaxies from \citet{reinesetal2013} are shown as pink points \citep[including NGC 4395;][]{filippenkosargent1989}. The dwarf galaxy RGG 118  \citep{reinesetal2013} hosting a $\sim$50,000 \msun\ BH \citep{baldassareetal2015} is the dark green point, and Pox 52 \citep{barthetal2004,thorntonetal2008} is the light green point (see \S\ref{sec:dwarfs}).  Fifteen reverberation-mapped AGN with BH masses taken from \citet{bentzkatz2015} are shown as purple points (see \S\ref{sec:rm}).  Dynamical BH mass measurements are taken from \citet{2013ARAA..51..511K} and shown as blue (elliptical galaxies), turquoise (S/S0 galaxies with classical bulges) and orange (S/S0 galaxies with pseudobulges) points.  The gray error bar indicates uncertainties in stellar masses for all points, and single-epoch spectroscopic BH masses. The gray lines show various \mbh\ vs. $M_{\rm bulge}$ relations based on ellipticals and spiral bulges with dynamical BH mass measurements.  The \citet{2013ARAA..51..511K} ``scaled" relation has bulge masses scaled down by 0.33 dex to account for differences in our assumed mass-to-light ratios (see \S\ref{sec:KHmasses}).  
\label{fig:mbhmstar}}
\end{figure*}

\citet{2013ARAA..51..511K} provide a different way to predict $M/L_K$ as a function of $B-V$ color (their equation 9).  This relation is based on the mass-to-light ratio calibrations of \citet{intoportinari2013}, shifted to a dynamically measurement zeropoint.  Using equation 9 in \citet{2013ARAA..51..511K} yields stellar masses that are systematically higher than the \citet{zibettietal2009} masses by 0.33 dex (see Figure \ref{fig:KHmasses}).  {This discrepancy is primarily due to different assumed stellar IMFs in the models and the shift of 0.1258 dex in log $M/L_K$ applied by \citet{2013ARAA..51..511K}.  \citet{zibettietal2009} adopt a Chabrier IMF, whereas \citet{intoportinari2013} assume a Kroupa IMF.   The stellar IMF is known to significantly affect the overall normalization of log $M/L_K$ \citep[e.g.,][]{belldejong2001}.}  Here we adopt the \citet{zibettietal2009} masses for consistency with the other samples in this work.

\section{Local BH Mass $-$ Total Stellar Mass Relations}\label{sec:relation}

One of the most useful aspects of BH-galaxy scaling relationships is that they provide a way to estimate BH mass from more easily measured galaxy properties.  While the tightest scaling relations appear to be between the BH mass and bulge properties in quiescent early-type galaxies, we seek to quantify the relationship between BH mass and total stellar mass to facilitate work at higher redshifts where measuring bulge properties is difficult or impossible and BHs are identified via nuclear activity.  

In Figure \ref{fig:mbhmstar} we plot BH mass versus host galaxy total stellar mass for our local sample of 244 broad-line AGN and the additional objects described in \S\ref{sec:add}.  A single linear relation is disfavored by the data.  There is a large range in BH mass at a given total stellar mass (e.g., a factor of $\sim$1000 in $M_{\rm BH}$ at $M_{\rm stellar} \sim 10^{10.5}~M_\odot$).  

Despite a significant amount of scatter in this plot, it is clear that at a given total stellar mass, AGN host galaxies at $z \sim 0$ tend to fall below elliptical galaxies and spiral/S0 galaxies with classical bulges hosting quiescent BHs.  {If the spectroscopic BH masses were shifted down (see \S\ref{sec:rm}), this would cause an even larger discrepancy between the AGNs and the dynamically detected BHs in Figure \ref{fig:mbhmstar}.}  Similarly, the AGN host galaxies fall below the canonical BH-to-bulge mass relations defined by these inactive early-type galaxies \citep[e.g.,][]{Haring2004,McConnell2013,2013ARAA..51..511K}.  BHs in galaxies with pseudobulges also preferentially lie below the scaling relations based on ellipticals and classical bulges \citep[also see][]{greeneetal2010,kormendyetal2011}.

We note that bulge mass and total stellar mass are equivalent for elliptical galaxies, which dominate the samples used to derive BH-to-bulge mass relations.  Indeed, we find that a linear regression using total stellar mass for the elliptical galaxies and spiral/S0 galaxies with classical bulges, which have dynamically-measured BH masses (\S\ref{sec:KHmasses}), is roughly consistent with standard bulge mass relations, albeit with more scatter (see below).

\begin{figure*}[!t]
$\begin{array}{cc}
{{\includegraphics[height=3.2in]{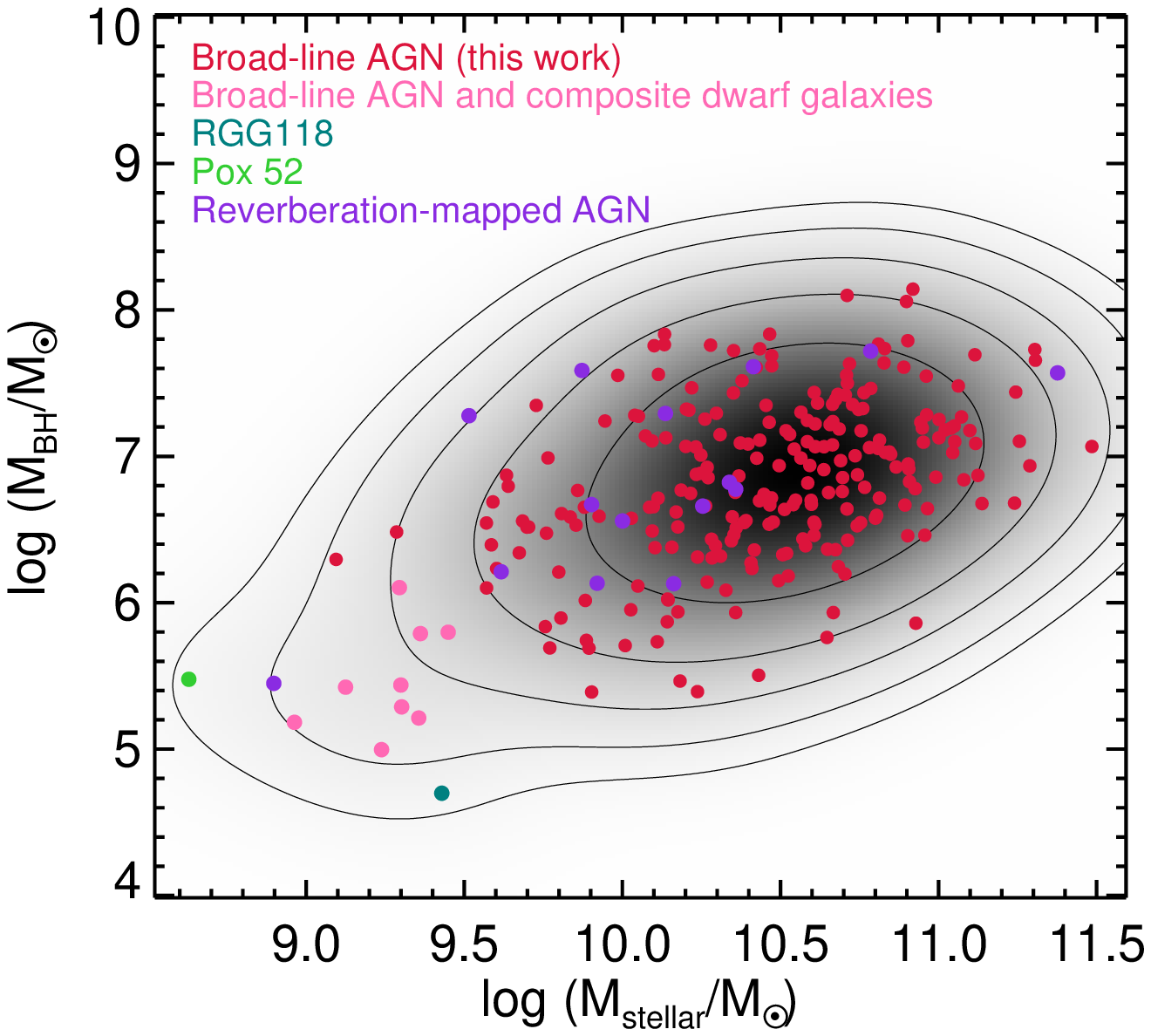}}} &
\hspace{-0.1cm}
{{\includegraphics[height=3.2in]{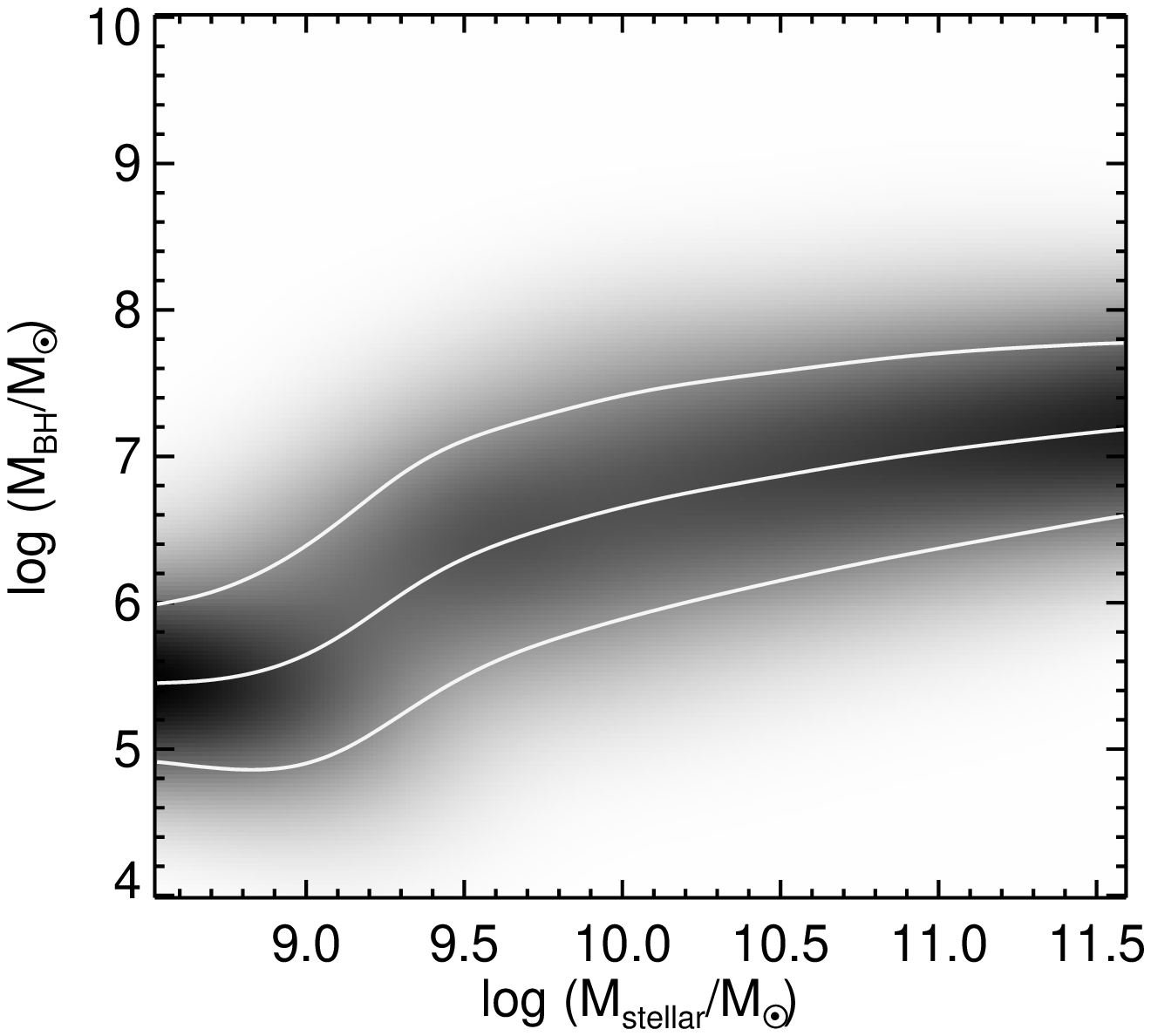}}}
\end{array}$
\caption{\footnotesize Left: Black hole mass versus total host galaxy stellar mass for local AGNs with the kernel density estimate (see text) shown in grayscale. Contour levels are at $(1/2)^n \times$ the peak value, where n = 1 to 5. 
Right: Conditional PDF $p({\rm log}M_{\rm BH} | {\rm log}M_{\rm stellar})$ computed by normalizing the kernel density estimate at each log $M_{\rm stellar}$. 
The middle line indicates the median of the PDF as a function of log~$M_{\rm stellar}$ and the outer white lines show the standard deviation.
\label{fig:kde}}
\end{figure*}

Thus, it appears that a separation exists between our sample of uniformly selected AGN hosts (\S\ref{sec:sample}), and ellipticals and classical bulges.   We anticipate that using or extrapolating the canonical BH-to-bulge mass scaling relations to interpret samples of galaxies with uncertain morphological classification, or AGN hosts, may lead to erroneous inferences.

\subsection{The BH-to-Total Stellar Mass Relation for Local AGNs}

We plot log $M_{\rm BH}$ versus log $M_{\rm stellar}$ for the AGNs alone in the left panel of Figure \ref{fig:kde}.  We first use a non-parametric method to help visualize the data and demonstrate that there is indeed a correlation between BH mass and total stellar mass for local AGNs.  We use the kernel density estimation technique \citep[e.g.,][]{Silverman1986} to estimate the density function in the log~$M_{\rm stellar} - {\rm log}~M_{\rm BH}$ plane from the observed data for all AGNs\footnote{For individual AGN with multiple BH mass estimates, we include only one data point with priority given to reverberation masses when available (e.g., NGC 4395).}.  Each data point is represented by a two-dimensional normalized Gaussian kernel.  The smoothing parameter (e.g., $\sigma$ for a Gaussian) is set to 0.3 and 0.5 for log~$M_{\rm stellar}$ and log~$M_{\rm BH}$, respectively, and reflects the measurement uncertainties for the majority of our sample (where masses are in units of $M_\odot$).  The individual kernels are then summed to produce the kernel density estimate (left panel of Figure \ref{fig:kde}).  The kernel density estimate is subsequently normalized for each log $M_{\rm stellar}$ independently to construct the conditional probability distribution function (PDF), $p({\rm log}M_{\rm BH} | {\rm log}M_{\rm stellar})$, which illustrates the dependence of BH mass on total stellar mass for our sample of AGNs.  The right panel of Figure \ref{fig:kde} shows the resulting PDF, where the lines correspond to the median and standard deviation as a function of log~$M_{\rm stellar}$.  

This non-parametric method nicely illustrates a correlation between log $M_{\rm BH}$ and log $M_{\rm stellar}$ for our sample of AGNs.  However, the shape/slope of the relation may be different for the {\it population} of local AGNs since the conditional PDF is based on data that is susceptible to selection biases that are particularly severe at low masses (see \S\ref{sec:dwarfs}).  While we should not immediately assume these data are well described by a linear relation, the sample Pearson correlation coefficient indicates that a linear relationship between log~$M_{\rm BH}$ and log~$M_{\rm stellar}$ is a reasonable description of the data; $r =  0.54$ with a probability $p<10^{-6}$ that no linear correlation is present.  We therefore use a line to parameterize the AGN relation. 

We take a Bayesian approach to linear regression using the method of \citet{kelly2007}\footnote{{\tt linmix\_err.pro} in the IDL Astronomy User's Library.}, which accounts for uncertainties in both log~$M_{\rm BH}$ and log~$M_{\rm stellar}$.  To facilitate comparison with other studies, we parameterize the relation as 
\begin{equation}
{\rm log}(M_{\rm BH}/M_\odot) = \alpha + \beta~{\rm log} (M_{\rm stellar}/10^{11} M_\odot)
\label{eqn:mbhmstar}
\end{equation}
\noindent
and find 
\begin{equation}
\alpha = 7.45 \pm 0.08; \beta = 1.05 \pm 0.11.  
\label{eqn:coef_agn}
\end{equation}

\begin{figure*}[!t]
$\begin{array}{cc}
{{\includegraphics[height=3.2in]{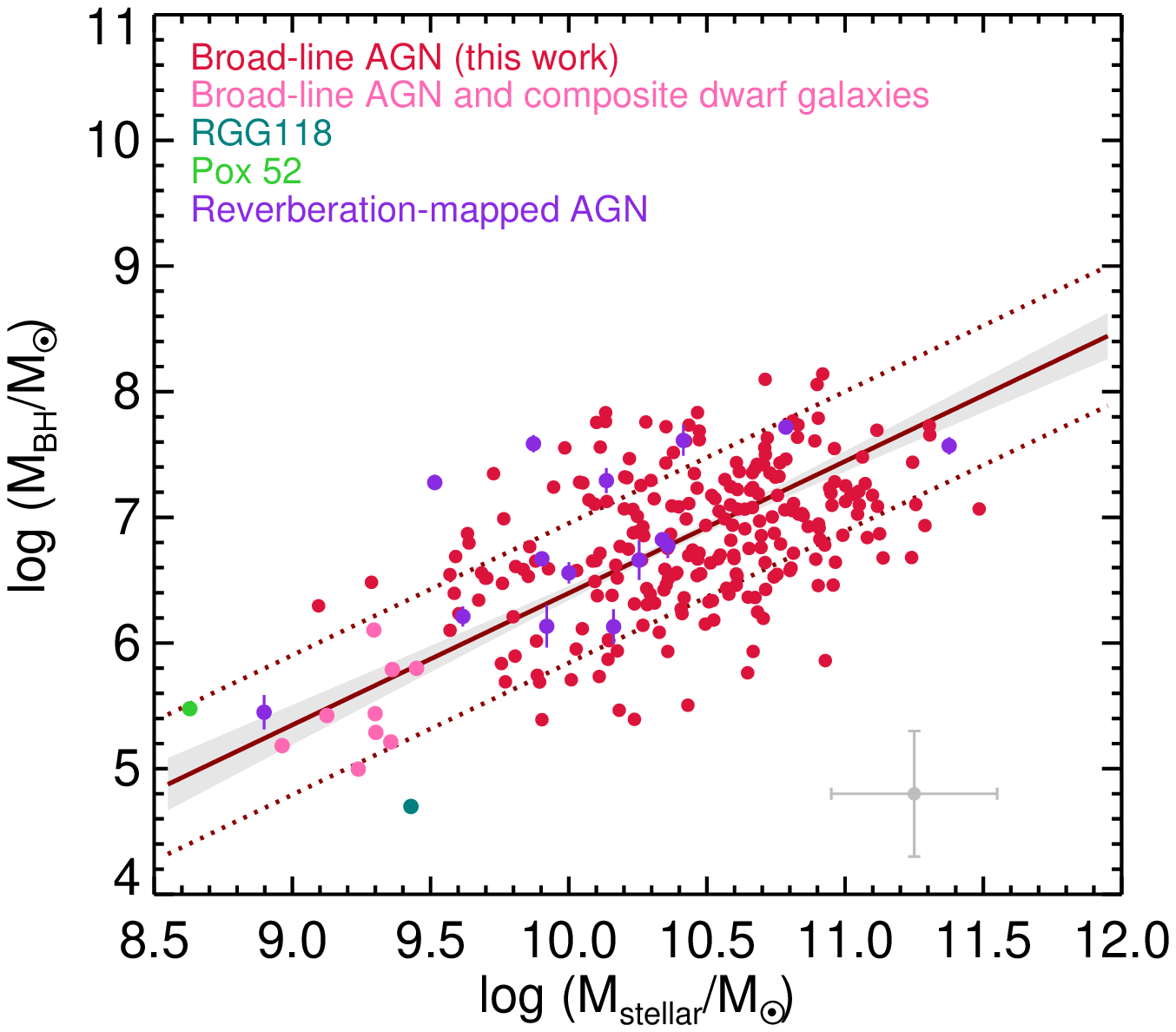}}} &
\hspace{-0.1cm}
{{\includegraphics[height=3.2in]{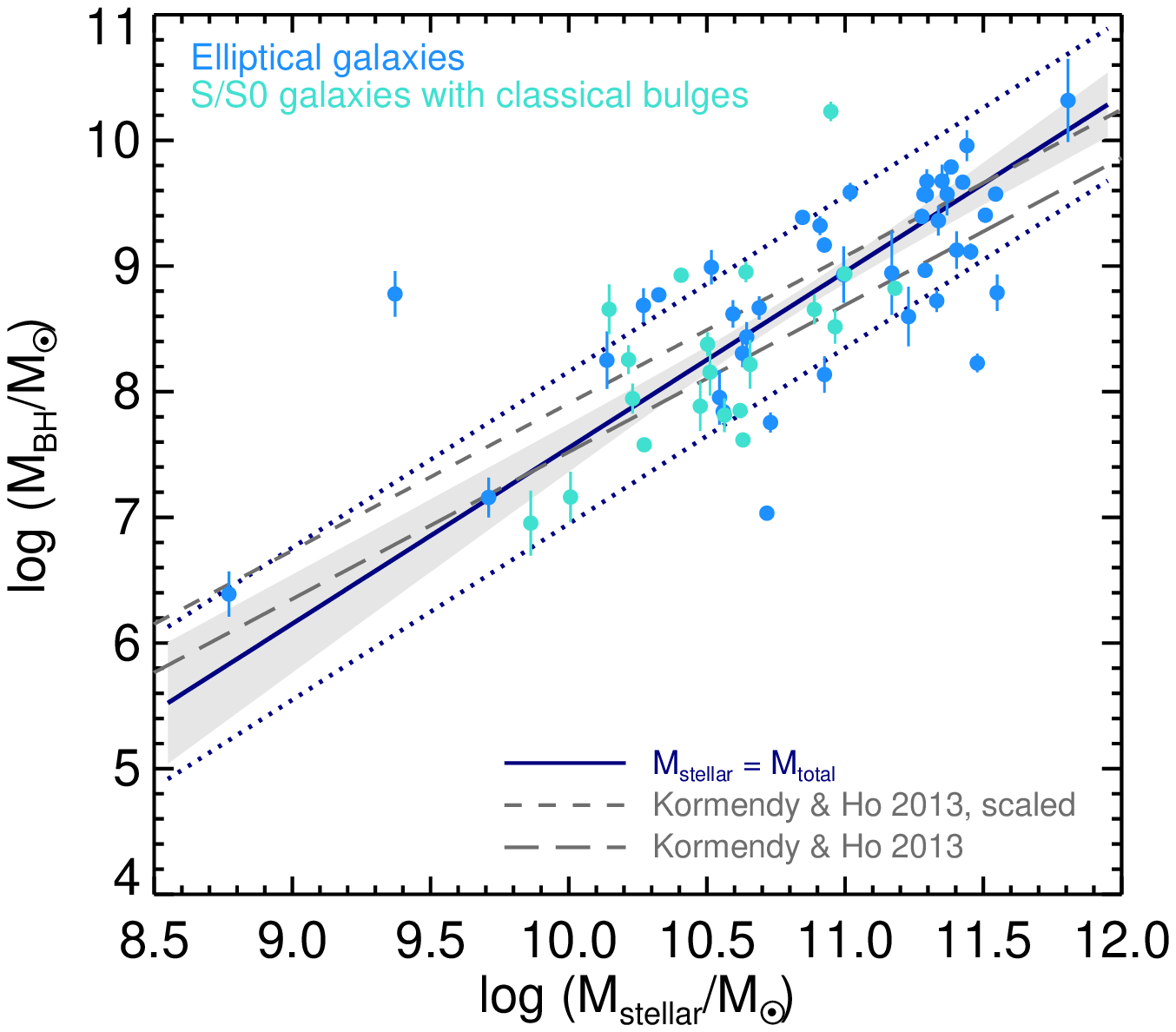}}}
\end{array}$
\caption{\footnotesize Left: The BH-to-total stellar mass relation for local AGNs (dark red line; equations \ref{eqn:mbhmstar} and \ref{eqn:coef_agn}).  The light gray shaded region 
accounts for the errors in the slope and intercept of the relation and the dark red dotted lines indicate the rms scatter of points around the relation (0.55 dex).  The gray error bar indicates uncertainties in stellar masses for all points, and single-epoch spectroscopic BH masses.  BH mass errors for the reverberation-mapped AGN are shown on the (purple) data points. 
Right: Same as the left panel, but for the inactive sample of elliptical galaxies and sprial/S0 galaxies with classical bulges (equations \ref{eqn:mbhmstar} and \ref{eqn:coef_eb}).  The dark blue line indicates our relation derived using total stellar mass (\S\ref{sec:KHmasses}).  
\label{fig:agnseq}}
\end{figure*}

\noindent
The quoted slope and intercept are given by the median of 10,000 draws from the posterior probability distribution of the parameters. The errors on the linear coefficients are correlated and the reported values are determined from the $1\sigma$ error ellipse. 
The rms deviation of the BH mass measurements from the relation is 0.55 dex, and incorporates both our adopted measurement errors of 0.50 dex and a best-fit intrinsic scatter of 0.24 dex (added in quadrature).  The intrinsic scatter may be larger if our measurement errors are overestimated.  The linear relation for the AGN host galaxies is shown in the left panel of Figure \ref{fig:agnseq}.  {We note that our results do not change significantly when using only our primary sample of broad-line AGNs (\S\ref{sec:sample}) and excluding the additional AGNs described in sections \ref{sec:dwarfs} and \ref{sec:rm}.  The slope and intercept derived from our uniformly selected sample agree with those in equation \ref{eqn:coef_agn} within the $1\sigma$ uncertainties.}

The BH-to-total stellar mass relation for the ellipticals and classical bulges is shown in the right panel of Figure \ref{fig:agnseq} for comparison.  The corresponding coefficients in equation \ref{eqn:mbhmstar} are given by    

\begin{equation}
\alpha = 8.95 \pm 0.09; \beta = 1.40 \pm 0.21,  
\label{eqn:coef_eb}
\end{equation}

\noindent
and the intrinsic scatter is 0.47 dex (rms = 0.60 dex).

\subsection{Implications for BH-to-Stellar Mass Fractions}

\begin{figure}[!h]
\hspace{0.1cm}
{\includegraphics[width=3.3in]{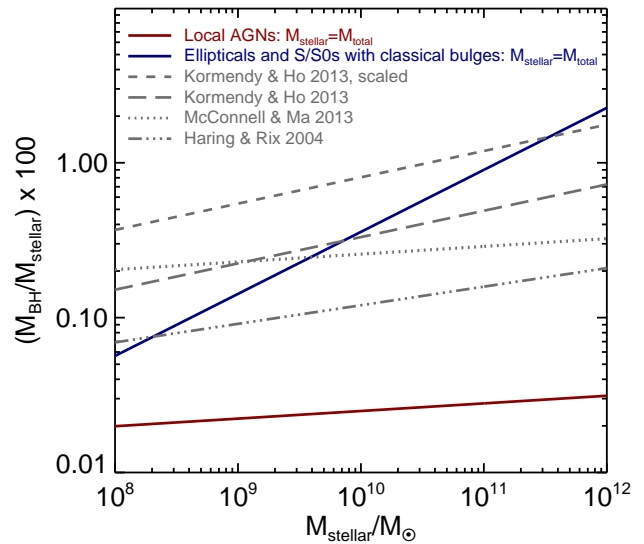}}
\caption{\footnotesize {BH mass fractions (given as a percentage of the stellar mass) as a function of stellar mass.  Our local AGN relation, where $M_{\rm stellar} = M_{\rm total}$ is shown as a dark red line.   Our total stellar mass relation for quiescent BHs in early-type galaxies is shown as a dark blue line.  Bulge mass relations from the literature are shown as gray lines.}   
\label{fig:mbh_frac}}
\end{figure}

Figure \ref{fig:mbh_frac} shows BH mass fractions as a function of stellar mass for our two {\it total} stellar mass relations at $z \sim 0$ (AGNs and dynamically measured BHs; equations \ref{eqn:mbhmstar}-\ref{eqn:coef_eb}), as well as some standard BH-to-{\it bulge} mass relations \citep[e.g.,][]{2013ARAA..51..511K, McConnell2013, Haring2004}.  The BH-to-total stellar mass fraction given by the AGN relation is $M_{\rm BH}/M_{\rm stellar} \sim 0.02\%$ to $0.03\%$ across the stellar mass range $10^8 \leq M_{\rm stellar}/M_\odot \leq 10^{12}$.  This is markedly smaller, by roughly an order of magnitude, than the BH-to-bulge mass fractions derived from quiescent early-type galaxies that are commonly used as references.

For instance, the BH mass fraction given by our AGN total stellar mass relation is $\sim 19$ to 56 times lower than that given by the canonical BH-to-bulge mass relation of \citet{2013ARAA..51..511K} across the stellar mass range $10^8 \leq M_{\rm stellar}/M_\odot \leq 10^{12}$, accounting for differences in our assumed stellar mass-to-light ratios (see \S\ref{sec:KHmasses}).  Using total stellar mass rather than bulge mass for the host galaxies of dynamically measured BHs in ellipticals and classical bulges also results in a BH mass fraction that is roughly an order of magnitude larger than that of the AGN host galaxies.  We thus urge extreme caution when using the canonical BH-to-bulge mass scaling relations as a proxy for BH-to-total stellar masses since this may lead to a biased interpretation.

\subsection{Systematic Uncertainties}

Based on the $z \sim 0$ samples presented in this work, we have shown that AGNs occupy a region in the log~$M_{\rm stellar} - {\rm log}~M_{\rm BH}$ plane below that populated by ellipticals and classical bulges.  Here we consider systematic uncertainties in stellar and BH masses that may affect this empirical result.  We note that differences in distance scales are negligible.  The range of assumed $H_0$ across all samples represented in Figure \ref{fig:mbhmstar} varies from 70 to 71 km s$^{-1}$ Mpc$^{-1}$.  

We do not expect that systematic uncertainties in stellar masses between the samples shown in Figure \ref{fig:agnseq} can account for the two different relations.  First, all stellar masses have been estimated in the most consistent manner feasible.  That is, we use color dependent mass-to-light ratios provided by \citet{zibettietal2009} with either a combination of $g$ and $i$-band SDSS data (all AGN except Pox 52), or $B,V$, and $K$-band data (galaxies with dynamical BH masses and Pox 52).  As described in \S\ref{sec:mstar}, we have accounted for any AGN contribution when calculating stellar masses so we do not think AGN host galaxy masses are significantly overestimated.  {Moreover, in order to bring the sample of AGNs onto the upper relation by shifting their stellar masses, the stellar masses would need to be reduced by more than an order of magnitude (see Figure \ref{fig:mbhmstar}).}  

The virial BH masses estimated for our sample of broad-line AGNs are quite indirect and subject to various uncertainties.  For example, the broad-line region geometry and orientation certainly varies between objects \citep{kollatschny2003,bentzetal2009lamp,denneyetal2010,barthetal2011}, yet we apply a single geometric scaling factor since we do not have this information for the individual broad-line AGNs in our sample.  There is also the possibility that there are nongravitational contributions to the measured gas velocities \citep[e.g.,][]{krolik2001}, although this would lead to systematically overestimated BH masses.

The lower relation defined by the broad-line AGNs has a normalization that is $\sim$1.2 dex lower than the upper relation at $M_{\rm stellar} = 10^{10} M_\odot$.  Across our sample, uncertainties in the virial BH mass estimates are expected to be on the order of $\sim 0.5$ dex \citep[e.g.,][]{vestergaardpeterson2006,Shen2013}, which is considerably less than the offset in BH mass between the two relations.   {Any reasonable variation in the virial factor can also be ruled out as producing artificially low BH masses for the broad-line AGNs.  In order to get AGN BH masses to fall on the upper relation by changing the virial factor, $<f>$ would need to be $\gtrsim 40$ ($\epsilon \gtrsim 10$ in equation 1).}  Finally, we note that dynamically measured BHs in galaxies with pseudobulges, as well as the reverberation-mapped AGN, overlap our sample of broad-line AGN.  For all of these reasons, we conclude that uncertainties in virial BH masses alone are not artificially producing a lower relation in the log~$M_{\rm stellar} - {\rm log}~M_{\rm BH}$ plane for the AGN. 

{We note that we have discarded 9 objects from our sample of AGNs in which the luminosity of the AGN dominates the total integrated photometry (host + AGN).  Our motivation for this was to minimize unreliable stellar mass estimates.  For a fixed stellar mass and Eddington ratio, this could bias us against large $M_{\rm BH}/M_{\rm stellar}$.  However, given that only $\sim 3\%$ of the AGNs were removed from our sample, we do not think this has impacted our results in any appreciable way.}

\subsection{Possible Origins for the Separation Between AGN Hosts and Ellipticals/Classical Bulges}\label{sec:origins}

{We now turn our attention to possible origins for two BH-to-total stellar mass relations: one comprising AGNs, the other ellipticals and classical bulges, with pseudobulges predominantly overlapping the AGNs.  It is worth noting that at least some of the galaxies with pseudobulges are active, as they are selected as being masers.  Others, like the Milky Way, are obviously inactive.

Given that the lower relation is defined by AGNs and the upper relation is defined by quiescent BHs with dynamical BH mass measurements, it is reasonable to consider if nuclear activity (or lack thereof) may be partially responsible for the existence of two separate realtions.  On the one hand, there are reasons to question the importance of nuclear activity since the samples defining the two relations are each fraught with their own selection biases.  Dynamical BH mass measurements are severely biased towards nearby, massive and dense galaxies where the BH sphere of influence can be resolved \citep[e.g.,][]{vandenbosch2015}.  {\citet{gultekinetal2011} showed that dynamical BH masses are not biased high in the $M_{\rm BH}-\sigma_\star$ plane for very large galaxies ($<\sigma_\star> \sim 268$ km s$^{-1}$), however we are probing significantly lower masses.}  There is no reason that galaxies with quiescent BHs should not overlap with the AGN host galaxies in Figure \ref{fig:mbhmstar}; we just cannot detect such BHs.    One the other hand, we see in Figure \ref{fig:mbhmstar} that we do not detect AGNs with BH masses as large as the quiescent BHs at a given stellar mass, suggesting nuclear activity is an important factor.  If there were AGNs with such large BH masses, we should see them since they would be bright.  Apparently, the larger BHs (at a given stellar mass) are not shining as AGNs \citep[see also, e.g.,][]{2004ApJ...613..109H,Merloni2004}.

Another potential factor is differences in host galaxy properties.  A comparison of the Hubble types of the galaxies with dynamically detected BHs \citep[from][]{2013ARAA..51..511K} supports this notion.  For example, the classical bulges are mostly hosted by early-type S0 galaxies (16/20 have Hubble types of S0 or SB0), and they overlap with the elliptical galaxies.  Alternatively, the pseudobulges that tend to overlap with the AGN host galaxies are more commonly found in later-type spiral galaxies (2/17 are found in SB0 galaxies, the rest are in spirals).  

\begin{figure}[!t]
\hspace{0.1cm}
{\includegraphics[width=3.3in]{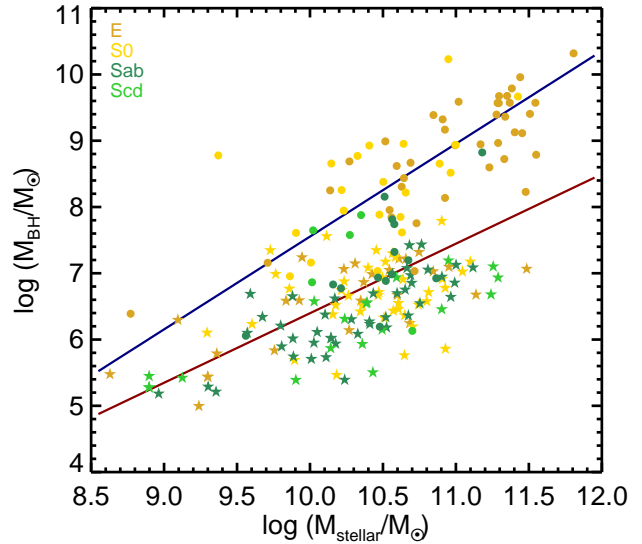}}
\caption{\footnotesize  Log~$M_{\rm BH}$ versus log~$M_{\rm stellar}$ with points color-coded by approximate morphological type (see text in \S\ref{sec:origins}).  {AGNs are plotted as stars and dynamically detected BHs are plotted as circles.}  The dark red line indicates our BH-to-total stellar mass relation based on local AGNs (equations \ref{eqn:mbhmstar} and \ref{eqn:coef_agn}).  The dark blue line indicates our BH-to-total stellar mass relation based on ellipticals and galaxies with classical bulges hosting dynamically detected BHs (equations \ref{eqn:mbhmstar} and \ref{eqn:coef_eb}).
\label{fig:hubbletype}}
\end{figure}

We find that a significant fraction of local broad-line AGN host galaxies are also spirals/disks, similar to what has previously been found in studies of moderate-luminosity AGN host galaxies out to $z \sim 3$ \citep[e.g., ][]{gaboretal2009,schawinskietal2011,kocevskietal2012,bennertetal2015}.  We obtained approximate Hubble types for a subset of the AGN host galaxies using automated morphological classifications from \citet{huertas-company2011}.  The automated classifications from this work have been shown to correlate well with visual classifications.  Rather than assigning a single morphological classification, \citet{huertas-company2011} provide probabilities of being in four morphological classes (E, S0, Sab, Scd).  For the AGN host galaxies with matches in the \citet{huertas-company2011} catalog (129 objects), we simply adopt the class with the highest probability.  

In Figure \ref{fig:hubbletype} we plot log~$M_{\rm BH}$ versus ${\rm log}~M_{\rm stellar}$ with points color-coded by morphological class.  We also include the galaxies with dynamical BH masses, converting Hubble types provided by \citet{2013ARAA..51..511K} to either E, S0, Sab, or Scd.  Based on these rough morphological classifications, it is clear that a significantly higher fraction of spiral galaxies lies close to the lower relation compared to the upper relation.  These spiral galaxies will have less prominent bulges than the galaxies on the upper relation, which are mostly bulge-dominated elliptical and S0 galaxies.  

{The AGN host galaxies on the lower BH-to-total stellar mass relation could conceivably follow the canonical BH-to-bulge mass relation\footnote{We have scaled the BH-to-bulge mass relation of \citet{2013ARAA..51..511K} to account for our different assumed stellar mass-to-light ratios as discussed in \S\ref{sec:KHmasses}.} if the classical bulge masses for the AGN hosts were on average only $\sim 5\%$ of the total stellar masses.  This echoes the result of the model for bulge evolution by \cite{2015ApJ...802..110L}, where their reference model (Model III) is a very good fit to our data.}  \cite{2014MNRAS.445.1261S} also find that low-mass quasars lie below the extrapolation of the local BH-to-bulge mass relation, but with a correction for the disk they obey it.  \citet{2014arXiv1411.3719C} advocate a lower $M_{\rm BH}/M_{\rm stellar}$ ratio for local active galaxies compared to inactive galaxies that have ``quenched" at earlier times.Ó

\section{Conclusions and Discussion}

In this paper we have studied the relation between BH mass and total stellar mass for nearby galaxies ($z<0.055$), including both galaxies with inactive BHs, with dynamical BH mass measurements, and galaxies with an active BH, with mass measurements based on reverberation mapping or single-epoch virial estimates.  Inclusion of the latter sources allows us to extend our sample to low BH masses, and also to use the same technique used at higher redshift.  Moreover, our stellar mass measurements rely on mass-to-light ratios, as is routinely the case for higher redshift samples. Therefore we build a local analog of higher redshift samples, but where we have better control on systematics. Our work is complementary to \cite{2014ApJ...780...70L}, where they measure in a uniform way the total luminosities of a sample of galaxies with dynamically-measured BHs, in that we try to provide a benchmark for observational or theoretical studies where detailed information on bulge properties is not available.  \citet{bennertetal2015} also took a similar approach towards the correlation with velocity dispersion.  One important caveat in our analysis is that the AGNs we have studied are moderate-luminosity Seyferts ($41.5 \lesssim {\rm log}~L_{\rm bol} \lesssim 44.4$).  At high-redshift, statistical samples are often biased towards more luminous AGNs and quasars. 

In Figure~\ref{fig:mbhmstar}, we plot BH mass versus total stellar mass for galaxies with $10^{8.5} \lesssim M_{\rm stellar} \lesssim 10^{12}~M_\odot$.  A single linear fit to the data is disfavored.  Rather, the AGN host galaxies define a relation that has a similar slope ($M_{\rm BH} \propto M_{\rm stellar}$) to early-type galaxies with quiescent BHs, but a normalization that is more than an order of magnitude lower (Figures \ref{fig:agnseq} and \ref{fig:mbh_frac}).  The different normalizations may be partially due to active versus inactive BHs, but can also be attributed to differences in host galaxy (bulge) properties as discussed in \S\ref{sec:origins}.  We caution that using the $z=0$ benchmark BH-to-bulge mass relations for AGN host galaxies, {or assuming $M_{\rm bulge}=M_{\rm total}$}, may lead to severely biased interpretations.  {For example, \citet{grahamscott2015} assume bulge mass equals total stellar mass for the 10 broad-line AGN and composite dwarf galaxies in \citet{reinesetal2013}, which in part leads them to conclude that AGN host galaxies follow a steeper, quadratic-like relation between BH mass and bulge mass.}  

Our work also has important implications for cosmological simulations that are tied to the local BH-to-bulge mass relations.
In most cosmological simulations that produce statistical samples of BHs and AGNs, i.e., large uniform-volume simulations (e.g., MassiveBlack I and II, \citet{dimatteoetal2012,degrafetal2014}; Illustris, \citet{sijackietal2014}; Eagle, \citet{schayeetal2015}; Horizon-AGN, \citet{duboisetal2014}) the resolution is limited to $\sim$kpc scales, making bulge-disk decomposition for low-mass galaxies unreliable. The typical approach, in fact, is to estimate the total stellar mass within the stellar half-mass radius \citep{sijackietal2014} or twice that \citep{degrafetal2014}, or extrapolate a fit to the mass profile of the
bulge inferred from kinematic data \citep{schayeetal2015}. If the simulations do not select ellipticals or galaxies with classical bulges and perform a bulge-disk decomposition, using the relations published in the literature for BH-bulge mass  \citep[e.g.,][]{Haring2004,MarconiHunt2003,McConnell2013,2013ARAA..51..511K} as a benchmark for the comparison between simulations and observations at $z=0$ (and perhaps beyond) would not be appropriate. Our method is easier to implement in the analysis of simulations, and can help disentangle issues related to BH growth and AGN properties.

Whether or not the BH-to-total stellar mass relation extends to even smaller masses bears directly to the origin of BH `seeds'. \cite{2009MNRAS.400.1911V} and \cite{2010MNRAS.408.1139V} suggested that if BH seeds are massive, e.g., $10^{5}$ \msun, as predicted by `direct collapse' models, the low-mass end of the relation between BHs and galaxies (they specifically referred to the correlation with the velocity dispersion, but the result would hold for the stellar mass as well) flattens towards an asymptotic value, creating a characteristic `plume' of ungrown BHs. Vice versa, if BH seeds are small, e.g., $10^{2}$ \msun, as predicted by models related to the first generation of stars, the expectation is that the observable scaling laws would not see the asymptotic value (the `plume') because it lies at masses below those that can be probed observationally.  

Finally, extending the sample at stellar masses $\sim 10^9$ \msun\ is fundamental to interpret results for much higher redshift galaxies with similar masses. Searches for AGN in galaxies with stellar masses $\sim 10^9$ \msun\ at $z>6$ have found very few, if any, BHs \citep{2011ApJ...742L...8W,fioreetal2012,2012ApJ...748...50C,2013ApJ...778..130T,2015arXiv150202562G,2015arXiv150106580W}.  If our sample is a good representation of the local universe, and the same relation based on local AGNs holds at high-redshift, it explains why we do not easily detect BHs at high-$z$: their masses would lie below the extrapolation of the local BH-bulge mass relation which is normally used as a benchmark. \cite{2011MNRAS.417.2085V} suggested that observations could indeed be explained with a BH-stellar mass correlation either steeper than at $z=0$, or with a lower normalization. \cite{2015arXiv150400018D} suggest that the growth of BHs in low-mass galaxies (galaxy mass $<10^{10}$ \msun\ and bulge mass  $<10^{9}$ \msun) is stunted because of supernova feedback,  which hinders accumulation of gas in the nucleus until the potential well of the bulge and galaxy become deep enough. In forthcoming work, we will explore the consequences of our results on the interpretation of high-$z$ BH populations, and the link to their hosts. 

\acknowledgements 
It is our pleasure to thank Rich Plotkin, Eric Bell, Kayhan G{\"u}ltekin and Jenny Greene for very helpful discussions. We also thank the anonymous referee for useful comments and suggestions. Support for AER was provided by NASA through Hubble Fellowship grant HST-HF2-51347.001-A awarded by the Space Telescope Science Institute, which is operated by the Association of Universities for Research in Astronomy, Inc., for NASA, under contract NAS 5-26555.
MV acknowledges funding from the European Research Council under the European 
Community's Seventh Framework Programme (FP7/2007-2013 Grant Agreement no.\ 614199, project ``BLACK'').  


\end{document}